\documentclass[11pt,preprint]{aastex}
\shorttitle{DIFFUSE INTERSTELLAR GAS TOWARD SN~2014J}
\shortauthors{RITCHEY ET AL.}

\begin{document}
\title{Diffuse Atomic and Molecular Gas in the Interstellar Medium of M82 toward SN~2014J}
\author{Adam M. Ritchey\altaffilmark{1}, Daniel E. Welty\altaffilmark{2}, Julie A. Dahlstrom\altaffilmark{3}, and Donald G. York\altaffilmark{2}$^,$\altaffilmark{4}}
\altaffiltext{1}{Department of Astronomy, University of Washington, Box 351580, Seattle, WA 98195; aritchey@astro.washington.edu}
\altaffiltext{2}{Department of Astronomy and Astrophysics, University of Chicago, 5640 S. Ellis Ave., Chicago, IL 60637}
\altaffiltext{3}{Department of Physics and Astronomy, Carthage College, 2001 Alford Park Dr., Kenosha, WI 53140}
\altaffiltext{4}{Also, Enrico Fermi Institute}

\begin{abstract}
We present a comprehensive analysis of interstellar absorption lines seen in moderately-high resolution, high signal-to-noise ratio optical spectra of SN~2014J in M82. Our observations were acquired over the course of six nights, covering the period from $\sim$6 days before to $\sim$30 days after the supernova reached its maximum $B$-band brightness. We examine complex absorption from Na~{\sc i}, Ca~{\sc ii}, K~{\sc i}, Ca~{\sc i}, CH$^+$, CH, and CN, arising primarily from diffuse gas in the interstellar medium (ISM) of M82. We detect Li~{\sc i} absorption over a range in velocity consistent with that exhibited by the strongest Na~{\sc i} and K~{\sc i} components associated with M82; this is the first detection of interstellar Li in a galaxy outside of the Local Group. There are no significant temporal variations in the absorption-line profiles over the 37 days sampled by our observations. The relative abundances of the various interstellar species detected reveal that the ISM of M82 probed by SN~2014J consists of a mixture of diffuse atomic and molecular clouds characterized by a wide range of physical/environmental conditions. Decreasing $N$(Na~{\sc i})/$N$(Ca~{\sc ii}) ratios and increasing $N$(Ca~{\sc i})/$N$(K~{\sc i}) ratios with increasing velocity are indicative of reduced depletion in the higher-velocity material. Significant component-to-component scatter in the $N$(Na~{\sc i})/$N$(Ca~{\sc ii}) and $N$(Ca~{\sc i})/$N$(Ca~{\sc ii}) ratios may be due to variations in the local ionization conditions. An apparent anti-correlation between the $N$(CH$^+$)/$N$(CH) and $N$(Ca~{\sc i})/$N$(Ca~{\sc ii}) ratios can be understood in terms of an opposite dependence on gas density and radiation field strength, while the overall high CH$^+$ abundance may be indicative of enhanced turbulence in the ISM of M82. The Li abundance also seems to be enhanced in M82, which supports the conclusions of recent gamma-ray emission studies that the cosmic-ray acceleration processes are greatly enhanced in this starburst galaxy.
\end{abstract}

\keywords{galaxies: individual (M82) --- galaxies: ISM --- ISM: abundances --- ISM: atoms --- ISM: molecules --- supernovae: individual (SN~2014J)}

\section{INTRODUCTION}
Bright extragalactic supernovae (SNe) provide unparalleled opportunities to study interstellar gas in nearby galaxies through high-resolution absorption-line spectroscopy. With the advent of modern, wide-field transient surveys, which quickly identify and classify newly-occurring supernovae, such studies are becoming increasingly more common. Still, the occurrence of a supernova as close as that recently discovered in M82 (SN~2014J; Fossey et al.~2014), especially one that probes as much interstellar material as this object does (e.g., Cox et al.~2014; Polshaw et al.~2014; Ritchey et al.~2014), is very rare. At a distance to M82 of approximately 3.5~Mpc (Jacobs et al.~2009; Dalcanton et al.~2009), SN~2014J is one of the closest SNe to be discovered in recent decades. Early photometric and spectroscopic analyses (Cao et al.~2014; Zheng et al.~2014; Goobar et al.~2014) indicate that SN~2014J is a normal Type~Ia supernova, but one that is heavily reddened by intervening interstellar material in the host galaxy. Estimates for the host galaxy contribution to $E(\bv)$ vary, but most are between about 0.8 and 1.3 mag (e.g., Welty et al.~2014; Amanullah et al.~2014; Foley et al.~2014; Marion et al.~2015). The proximity of SN~2014J makes it possible to obtain high-resolution, high signal-to-noise ratio (S/N) optical spectra at multiple epochs, while the significant dust extinction means that many interstellar species are likely to be detected. Such data can be used to search for weak interstellar features rarely seen in extragalactic sight lines and to look for temporal changes in the absorption profiles that might signify an interaction between the supernova and possible circumstellar material associated with the progenitor system (e.g., Patat et al.~2007; Simon et al.~2009) or, alternatively, could be used to probe small-scale structure in the interstellar medium (ISM) of the host galaxy (Patat et al.~2010).

Most optical studies of interstellar lines in supernova spectra must necessarily rely on the strong Na~{\sc i} and Ca~{\sc ii} features, as these are frequently the only lines detectable at the host galaxy velocity. The Na~{\sc i} and Ca~{\sc ii} profiles are useful for examining the basic kinematics of the absorption systems, and the Na~{\sc i}/Ca~{\sc ii} column density ratio can give a rough indication of the overall physical conditions (although there is some ambiguity between the effects of dust depletion, which mainly affects the Ca~{\sc ii} abundance, and ionization, which can influence the amount of Na~{\sc i} and Ca~{\sc ii} present). However, in situations where additional interstellar species are detectable, such as K~{\sc i}, Ca~{\sc i}, CH$^+$, CH, and CN, a more complete picture of the host galaxy ISM can be constructed. While such detections are rare, there have been a few SN absorption-line studies where these additional species were observed. D'Odorico et al.~(1989) presented moderately-high resolution optical spectra of the reddened Type Ia SN~1986G in NGC~5128 (Cen A); their observations yielded the first detections of interstellar CH and CH$^+$ in a galaxy outside of the Local Group. Vladilo et al.~(1994), examining high-resolution optical spectra of the Type IIb SN 1993J in M81, were the first to detect K~{\sc i} and Ca~{\sc i} absorption associated with gas beyond the Local Group. In this case, the absorbing material was plausibly identified as tidal debris resulting from interactions between M81 and M82 (see also Bowen et al.~1994; Roth et al.~2000; Lauroesch et al.~2005). Several groups obtained high-resolution spectra of the heavily-reddened Type Ia SN~2006X in M100, which exhibited extraordinarily strong host-galaxy absorption from interstellar CN, along with more moderate absorption from CH, CH$^+$, Ca~{\sc i}, and K~{\sc i} (Lauroesch et al.~2006; Patat et al.~2007; Cox \& Patat 2008; Phillips et al.~2013). Cox \& Patat (2014) recently presented high-resolution optical spectra of the reddened Type Ia SN~2008fp, which also showed strong CN absorption lines at the velocity of the host galaxy, ESO~428-G14, and even revealed host-galaxy absorption features of C$_2$ and C$_3$, which had previously not been detected in a galaxy beyond the Local Group.

In this paper, we present multi-epoch, moderately-high resolution, high S/N optical spectra of SN~2014J obtained with the ARC echelle spectrograph (ARCES; Wang et al.~2003) at Apache Point Observatory (APO). Our extensive, high-quality data set allows us to perform a comprehensive analysis of the interstellar absorption features arising from Milky Way and M82 gas in the direction of the supernova. Moreover, our observations permit us to make the first detection of Li~{\sc i} in a galaxy outside of the Local Group, from which we derive an estimate for the Li abundance in the ISM of M82. We describe the observations and the procedures used for data reduction and analysis in Section~2. In Section~3, we present the basic results concerning the column densities and component structure of the observed atomic and molecular species. In Section~4.1, we examine the kinematics of the absorption systems seen along the line of sight, while, in Section~4.2, we evaluate the physical conditions in the Milky Way and M82 components. In Section~4.3, we present a comparison between the closely-spaced sight lines to SN~1993J in M81 and SN~2014J in M82, and, in Section~4.4, we discuss the interstellar Li abundance in M82. We summarize our results and conclusions in Section~5. Two companion papers focus on a comparison between several of the prominent diffuse interstellar bands (DIBs) and the known atomic and molecular species observed in the ISM of M82 (Welty et al.~2014), and on a more complete census of the DIBs detected toward SN~2014J (D.~York et al., in preparation).

\section{OBSERVATIONS AND DATA ANALYSIS}
We obtained 47 individual exposures of SN~2014J with ARCES over the course of six nights (2014 January 27, 30; February 8, 11, 22; March 4), which cover the period from $\sim$6 days before to $\sim$30 days after the supernova reached its maximum $B$-band brightness on 2014 February 2 (Foley et al.~2014). Individual exposure times were limited to 20 minutes in order to minimize the number of cosmic-ray hits occurring during each integration. The echelle spectrograph provides complete wavelength coverage in the range 3800--10200~\AA{} with a resolving power of $R\approx31,500$ ($\Delta v\approx9.5$ km s$^{-1}$). Two of us (A.R. and J.D.) independently reduced the raw spectral images, following standard procedures for ARCES data (e.g., Thorburn et al.~2003). The spectra produced by the two reduction procedures show very good agreement, with only minor differences, which may be attributed to slight differences in the specific parameters and/or techniques adopted during the cosmic-ray removal process. Corrections for telluric absorption were applied to the individual extracted spectra using contemporaneous observations of a bright, lightly-reddened standard star (e.g., $\eta$~UMa). The corrected spectra were then shifted to the reference frame of the local standard of rest (LSR), and the 5--13 individual spectra obtained on a given night were co-added to produce nightly sum spectra. Details concerning the nightly sums, including total exposure times and estimates for the S/N at various wavelengths can be found in Table~1.

\begin{figure}
\centering
\includegraphics[width=0.7\textwidth]{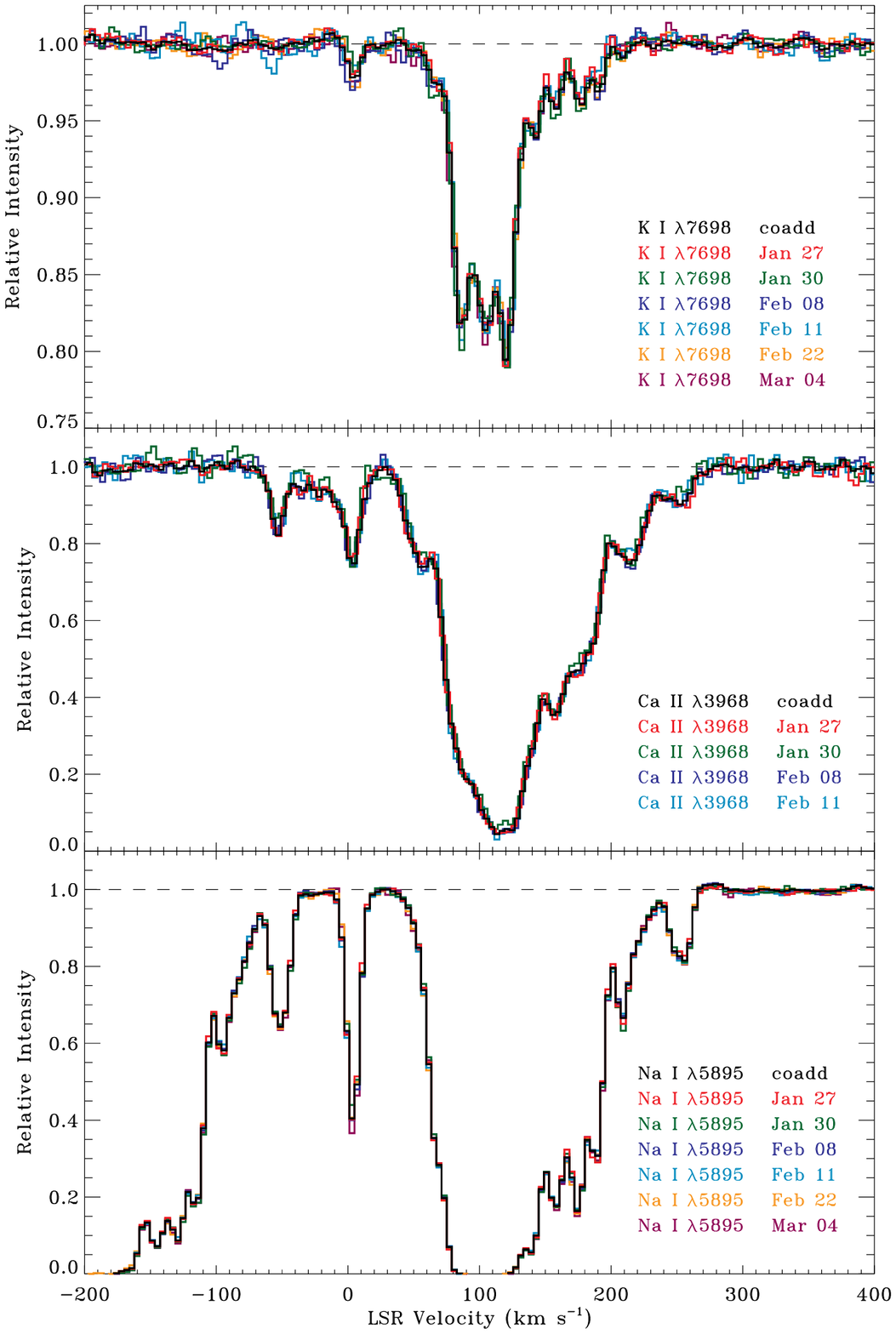}
\caption[]{Nightly sum spectra of the K~{\sc i}~$\lambda7698$, Ca~{\sc ii}~$\lambda3968$, and Na~{\sc i}~$\lambda5895$ absorption lines toward SN~2014J. There are no significant changes in these absorption features over the 37 days sampled by our observations (see also Figure~2). Only four epochs are shown for Ca~{\sc ii}~$\lambda3968$ because the supernova had very little blue flux on the remaining nights, making those Ca~{\sc ii} spectra unreliable.}
\end{figure}

\begin{figure}
\centering
\includegraphics[width=0.7\textwidth]{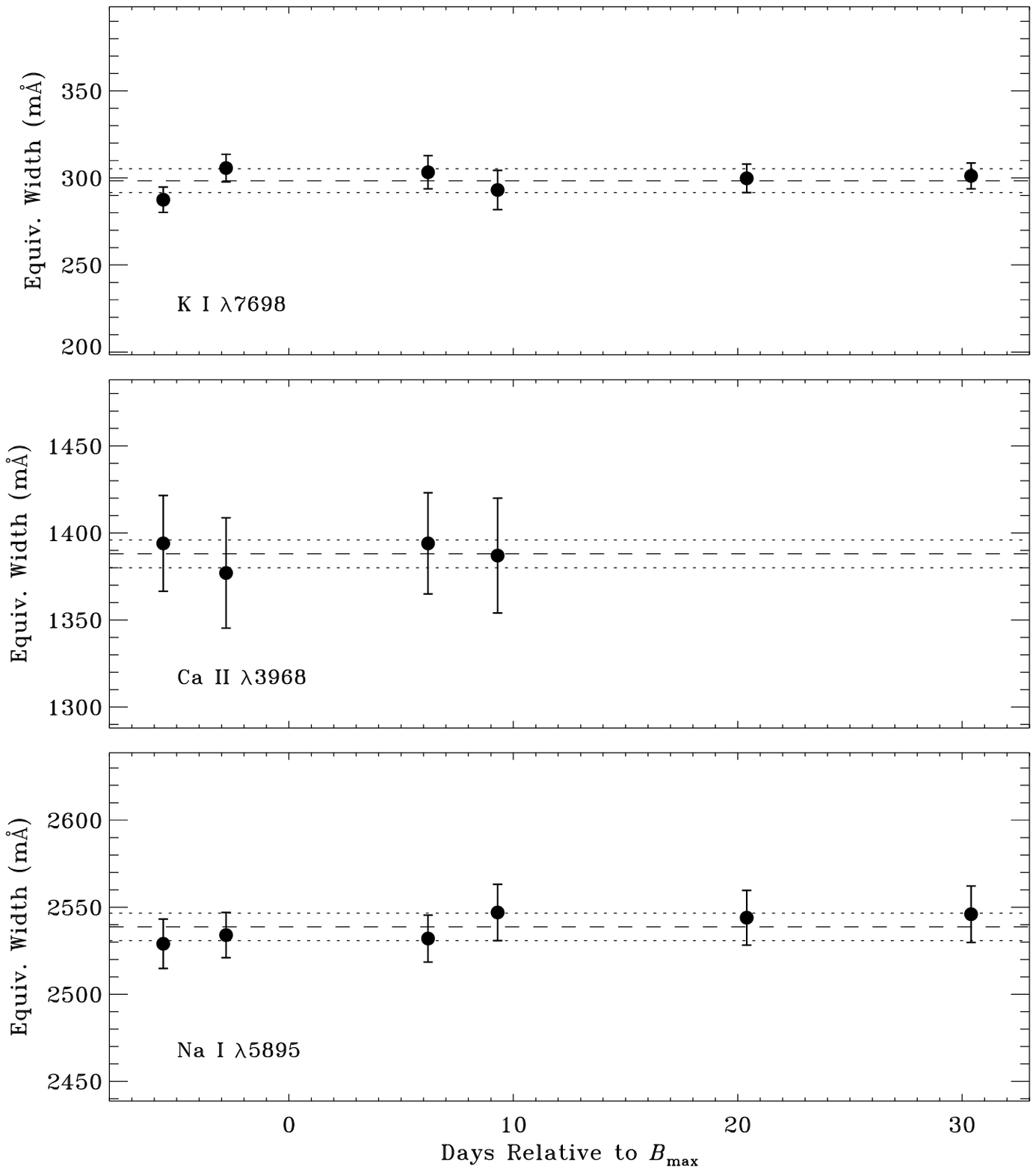}
\caption[]{Single-epoch equivalent widths of the K~{\sc i}~$\lambda7698$, Ca~{\sc ii}~$\lambda3968$, and Na~{\sc i}~$\lambda5895$ absorption features associated with M82 plotted as a function of light curve phase (see Table~1). The dashed and dotted lines give the mean and $\pm$1$\sigma$ standard deviation, respectively, for each set of measurements: $298.4\pm6.8$~m\AA{} for K~{\sc i}, $1388.0\pm8.0$~m\AA{} for Ca~{\sc ii}, and $2538.7\pm7.9$~m\AA{} for Na~{\sc i}. These mean values are consistent with the equivalent widths given in Table~2, which were computed directly from the co-added spectrum.}
\end{figure}

From a preliminary analysis of the spectra obtained on the first two nights, Ritchey et al.~(2014) described the extensive nature of the interstellar absorption features seen in the direction of SN~2014J. Absorption from Na~{\sc i}, Ca~{\sc ii}, K~{\sc i}, Ca~{\sc i}, CH$^+$, CH, and CN is clearly visible, and numerous DIBs are also present (Welty et al.~2014; D.~York et al., in preparation). In this paper, we focus mainly on the atomic and molecular species. We first examine the nightly sum spectra to determine if there are any temporal changes in the absorption-line profiles that could indicate the presence of circumstellar material associated with the progenitor of SN~2014J or could be used as a tracer of small-scale structure in the ISM of M82. However, as explained in more detail in Section~2.1, we do not find any evidence for temporal variations in the absorption profiles. Thus, our primary conclusions regarding the column densities and component structure of the interstellar species detected toward SN~2014J are based on our analysis of co-added spectra, as described in Section~2.2. The analysis of the Li~{\sc i} region is discussed in Section~2.3, as this part of the co-added spectrum required particular attention.

\subsection{Analysis of Nightly Sum Spectra}

Figure~1 displays the nightly sum spectra (normalized to the continuum) for K~{\sc i}~$\lambda7698$, Ca~{\sc ii}~$\lambda3968$, and Na~{\sc i}~$\lambda5895$. These lines are strong enough to be adequate tracers of circumstellar/interstellar material, yet are somewhat less affected by saturation than the stronger lines of the K~{\sc i}, Ca~{\sc ii}, and Na~{\sc i} doublets. As Figure~1 demonstrates, there are no significant changes in the absorption profiles of these lines over the 37 days sampled by our observations. (Only four epochs are shown for Ca~{\sc ii}~$\lambda3968$ because the supernova had very little blue flux on the remaining nights, making those Ca~{\sc ii} spectra unreliable; however, this does not affect our basic conclusion regarding the lack of variability.) In quantitative terms, the standard deviations of the single-epoch equivalent widths (computed only for absorption associated with M82; see Section~3) are smaller than the average uncertainties in those equivalent widths for the lines shown in Figure~1. The single-epoch equivalent widths are listed in Table~1, and are plotted as a function of light curve phase in Figure~2. The standard deviations for these equivalent width measurements are 6.8~m\AA{} for K~{\sc i}~$\lambda7698$, 8.0~m\AA{} for Ca~{\sc ii}~$\lambda3968$, and 7.9~m\AA{} for Na~{\sc i}~$\lambda5895$, while the average uncertainties in the measurements for the three lines are 8.6~m\AA{}, 30.3~m\AA{}, and 14.8~m\AA{}.

Our conclusion regarding the lack of temporal changes in the absorption profiles is consistent with that of Foley et al.~(2014), who saw no evidence of variation in the Na~{\sc i}~D lines, in the K~{\sc i}~$\lambda7664$ line, or in the $\lambda$5780.5 and $\lambda$5797.1 DIBs toward SN~2014J. From estimates of the S/N in our nightly sum spectra (for the stronger lines of the K~{\sc i}, Ca~{\sc ii}, and Na~{\sc i} doublets), we obtain 3$\sigma$ upper limits on the column densities of any undetected components of log~$N$(K~{\sc i})~$\lesssim$~10.2, log~$N$(Ca~{\sc ii})~$\lesssim$~11.0, and log~$N$(Na~{\sc i})~$\lesssim$~10.3. (Units for $N$ are cm$^{-2}$.) The limiting Na~{\sc i} column density is applicable for velocities in the range $-$500~$\lesssim~v_{\mathrm{LSR}}~\lesssim$~+1500~km~s$^{-1}$, while the Ca~{\sc ii} limit applies to velocities in the range $-$1000~$\lesssim~v_{\mathrm{LSR}}~\lesssim$~+1200~km~s$^{-1}$, as these were the regions typically examined for interstellar and/or circumstellar absorption features. Since any apparent differences in the K~{\sc i}, Ca~{\sc ii}, and Na~{\sc i} absorption profiles obtained at different epochs are consistent with the noise in the continuum (see Figure~1), these limiting column densities may also be interpreted as limits on the magnitude of any temporal variations. Fairly stringent upper limits on the density of circumstellar/interstellar material surrounding SN~2014J have already been derived from deep X-ray and radio observations (Margutti et al.~2014; P{\'e}rez-Torres et al.~2014). Our results provide additional support for the lack of any significant material in the immediate vicinity of the SN.

\subsection{Analysis of Co-added Spectra}

Having found no evidence for temporal variations in the absorption-line profiles, we co-added all of the 47 individual spectra to produce a single high S/N spectrum. (Below 4500~\AA{}, however, only the 29 spectra obtained on the first four nights were included in the sum due to the substantial decrease in the blue flux on the remaining two nights.) Portions of the spectra surrounding interstellar lines of interest were then normalized via low-order polynomial fits to the continuum regions. Equivalent widths for all of the observed atomic and molecular features (Table~2), along with apparent optical depth (AOD) estimates for the column densities, were determined from the normalized spectra by direct integration over the line profiles. Uncertainties in the measured equivalent widths include contributions from both photon noise and continuum placement (Jenkins et al.~1973; Sembach \& Savage 1992). Final column densities for the atomic and molecular species were derived through independent multi-component fits to the normalized absorption profiles (Figure~3), using the codes \emph{ismod} (A.R.; see Sheffer et al.~2008) and \emph{fits6p} (D.W.; see Welty et al.~2003). The profile fitting routines also yielded estimates for the line widths ($b$-values) and velocities of the absorption components, although ``individual'' interstellar clouds are most likely not resolved at the resolution of ARCES, and the $b$-values derived here are probably more accurately referred to as ``effective'' $b$-values.

\begin{figure}
\centering
\includegraphics[width=0.49\textwidth]{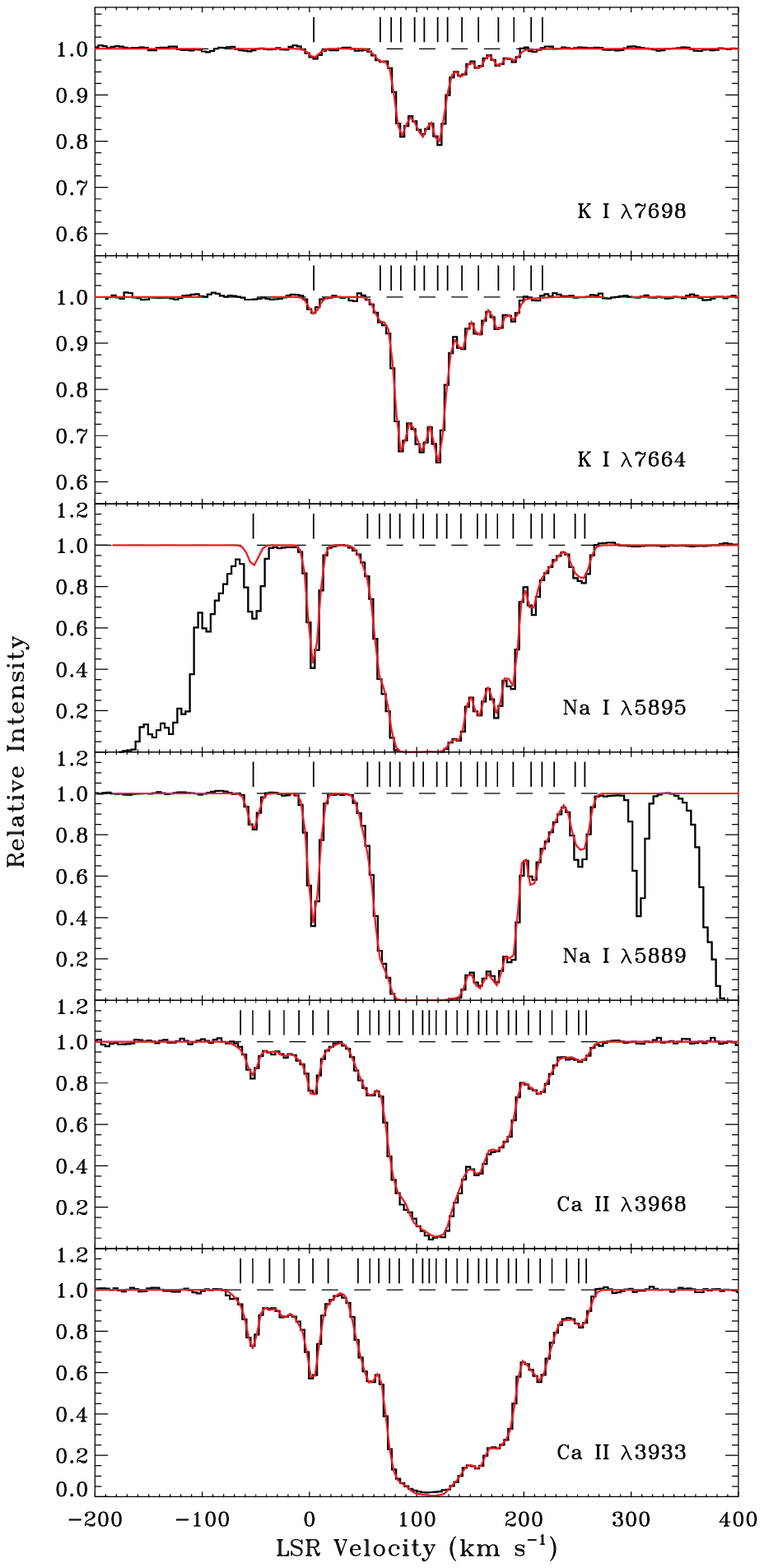}
\includegraphics[width=0.49\textwidth]{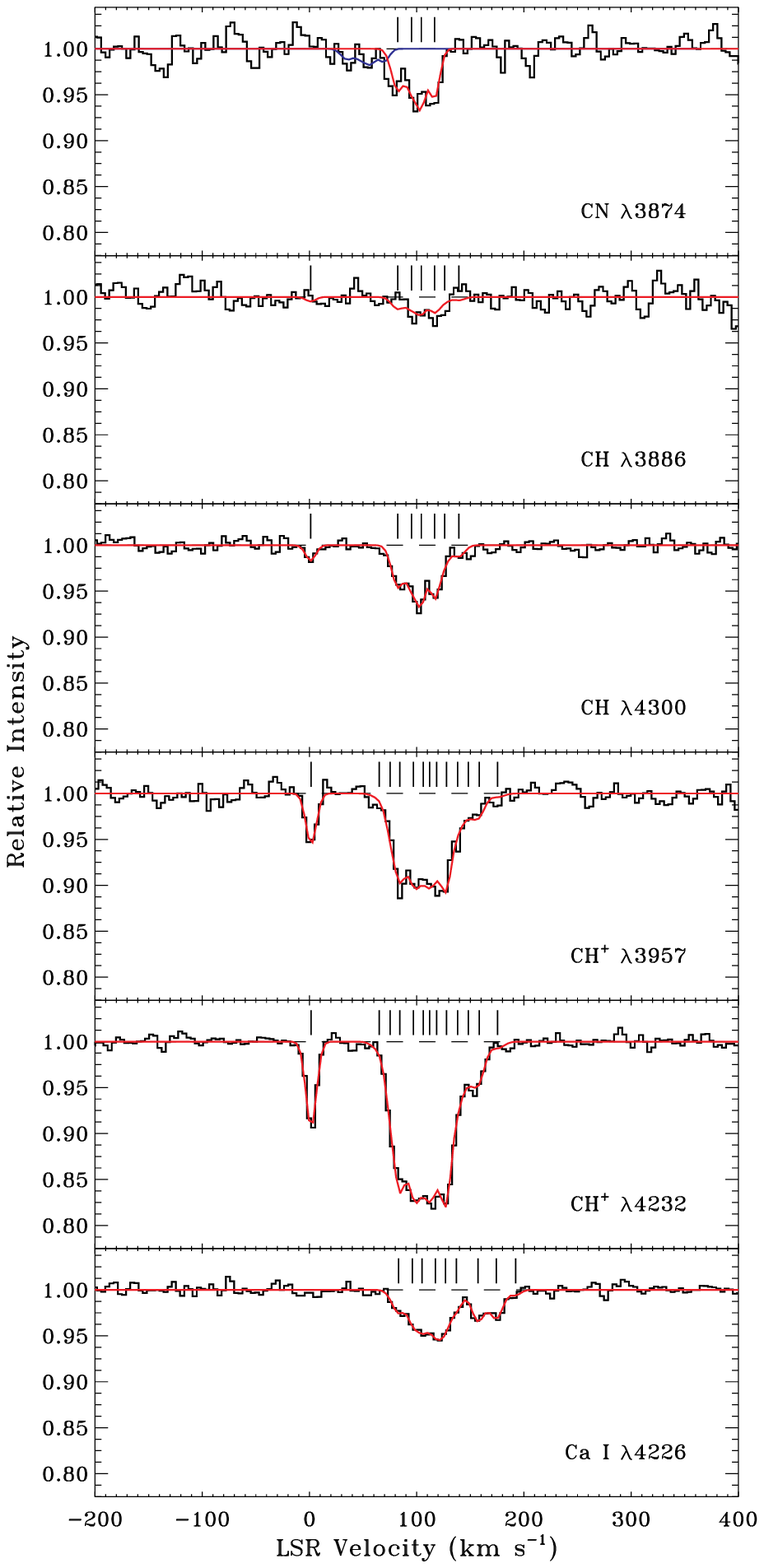}
\caption[]{Multi-component profile synthesis fits to the atomic and molecular lines observed in the direction of SN~2014J. Synthetic absorption profiles (colored curves) are shown superimposed onto the observed spectra (black histograms). Tick marks give the positions of the individual components included in the fits. (The synthetic profiles shown here correspond to the final component structures obtained for the various species, rather than to fits to the individual lines.) For CN, the red curve (with tick marks) shows the fit to the $R$(0) line, while the blue curve is for $R$(1). Note that the Na~{\sc i}~D$_1$ component at $-$52~km~s$^{-1}$ is blended with D$_2$ absorption near +252~km~s$^{-1}$.}
\end{figure}

The K~{\sc i} profiles were the first to be analyzed, since these lines are of moderate strength and often have narrow line widths, making the individual components easily discernible. The K~{\sc i} component structure then served as a basis for fitting the other atomic and molecular lines, adding or removing components as necessary, allowing for small changes in the relative velocities (in some cases), and (generally) constraining the $b$-values to fall between 1.0 and 6.0~km~s$^{-1}$. Our overall aim was to fit the ``obvious'' features in the spectra (considering asymmetries and inflections within the profiles), while trying not to overfit the data (e.g., such that the root mean square (RMS) of the fit was less than the local continuum RMS). Fits to the Na~{\sc i} profiles required special attention for two reasons. First, the absorption profiles are strongly saturated for +80~$\lesssim~v_{\mathrm{LSR}}~\lesssim$~+120~km~s$^{-1}$. Thus, the Na~{\sc i} column densities for the four (assumed) components in this range were held fixed, assuming $N$(Na~{\sc i})/$N$(K~{\sc i})~$\approx$~90 (e.g., Welty \& Hobbs 2001). We note that $N$(Na~{\sc i})/$N$(K~{\sc i}) ratios much different from 90 for the strongest components would not be very consistent with the observed profiles or with the ratios seen for the weaker components. Second, the Na~{\sc i}~D$_1$ component at $-$52~km~s$^{-1}$ is blended with D$_2$ absorption near +252~km~s$^{-1}$ (see Figure~3). We therefore held the column density of the D$_1$ component at $-$52~km~s$^{-1}$ fixed based on the result for the D$_2$ component at the same velocity. The D$_2$ absorption near +252~km~s$^{-1}$ could then be fitted after removing the contribution from D$_1$. The CN $\lambda3874$ absorption profile also presented a challenge due to possible blending of the $R$(0) and $R$(1) lines (which are separated by $\sim$47~km~s$^{-1}$). In fitting these features, we adopted the component structure found for CH~$\lambda4300$, and held the absolute velocities of the absorption components fixed.

\subsection{Analysis of the Li~{\sc i} Region}

The significant column densities of Na~{\sc i} and K~{\sc i} probed by SN~2014J (see Table~3), and the very high S/N achieved in our total combined spectrum summed over all epochs ($\sim$1100 near 6700~\AA{}), indicated that detectable absorption from Li~{\sc i}~$\lambda6707$ might be present in the data. A portion of the combined spectrum in the vicinity of the Li~{\sc i} line is shown in the upper panel of Figure~4. The analysis of this region of the spectrum is complicated by the presence of weak DIBs, which are known from studies of heavily-reddened Galactic stars (e.g., HD~183143; Hobbs et al.~2009), and which are clearly detected toward SN~2014J. The locations of four of these DIBs, those with (rest-frame) wavelengths of 6702.0~\AA, 6706.6~\AA, 6709.4~\AA, and 6713.8~\AA, are indicated in Figure~4. Considering the spread in velocity of the strongest K~{\sc i} components seen toward SN~2014J, any associated Li~{\sc i} absorption is likely to be blended with absorption from the nearby $\lambda$6706.6 DIB. We therefore applied a method analagous to that described by Welty et al.~(2014) to model and remove the absorption from the DIBs in this region of the spectrum. A template for the ``intrinsic'' DIB profiles was constructed from ARCES spectra of HD~183143, which has a color excess similar to that of SN~2014J, and exhibits similar relative DIB strengths (Hobbs et al.~2009; Friedman et al.~2011; Welty et al.~2014). The template was then duplicated for each of the velocity components seen in K~{\sc i} toward the SN, weighting the contributions from the stronger main components and the weaker, higher-velocity components separately so as to provide an approximate fit to the observed DIB profiles (see Welty et al.~2014). In general, the fits are substantially improved if the weaker K~{\sc i} components are given a higher weight. In the adopted model DIB spectrum (see Figure~4), the weighting factors for the weaker components (relative to the main components) are one for the $\lambda$6702.0 DIB, seven for the $\lambda$6706.6 DIB, three for the $\lambda$6709.4 DIB, and two for the $\lambda$6713.8 DIB, consistent with the range of results found for the DIBs studied by Welty et al.~(2014).

\begin{figure}
\centering
\includegraphics[width=0.9\textwidth]{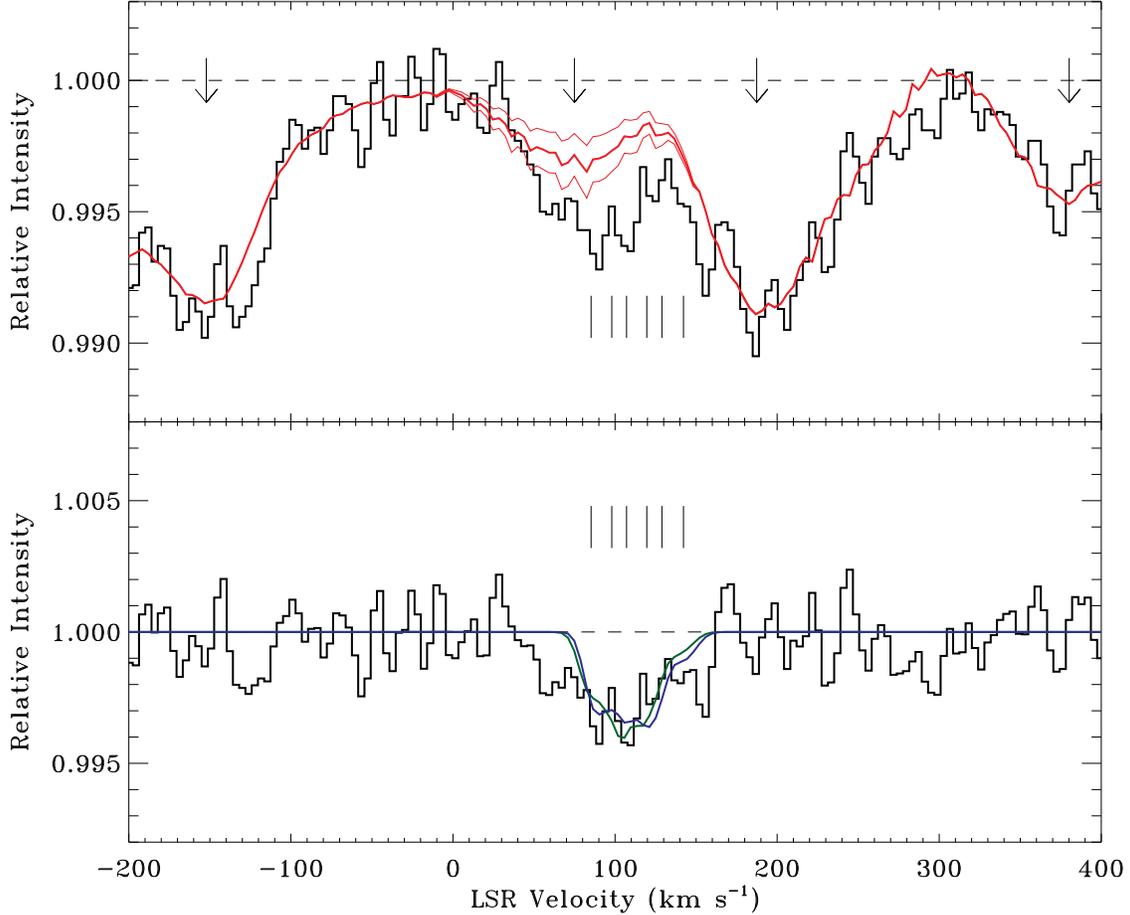}
\caption[]{Total combined spectra of SN~2014J in the vicinity of Li~{\sc i}~$\lambda6707$. The upper panel displays the continuum-normalized spectrum, which shows absorption from several weak DIBs surrounding the Li~{\sc i} feature. Arrows mark the positions of the DIBs with (rest-frame) wavelengths of 6702.0~\AA, 6706.6~\AA, 6709.4~\AA, and 6713.8~\AA. The thick red line in the upper panel represents the adopted model DIB spectrum, obtained using a technique analagous to that described by Welty et al.~(2014) (see Section~2.3). The thinner red lines represent the range of model spectra considered in fitting the $\lambda$6706.6 DIB, and show the effect of changing the overall scaling factor for $\lambda$6706.6 by $\pm2$ compared to the adopted model. The lower panel gives the residual spectrum after dividing by the adopted model spectrum. The blue and green lines in the lower panel represent profile synthesis fits to the Li~{\sc i} feature using templates based on K~{\sc i} and CH, respectively. Tick marks give the positions of the six components included in the profile template based on K~{\sc i}.}
\end{figure}

The adopted model DIB spectrum was divided into the observed spectrum, and the residual spectrum was then analyzed with a profile synthesis routine to derive the Li~{\sc i} column density. Two Li~{\sc i} syntheses were attempted, adopting the K~{\sc i} component structure for one and the CH component structure for the other (and using only the six components seen in both K~{\sc i} and CH for both syntheses). These three species are known to be well correlated in the local Galactic ISM (e.g., Welty \& Hobbs~2001; Knauth et al.~2003), and should therefore be distributed similarly along the line of sight. Indeed, the difference between the two profile templates used to fit the Li~{\sc i} absorption feature (lower panel of Figure~4) is minimal, and the Li~{\sc i} column densities resulting from the two syntheses are very similar: log~$N$(Li~{\sc i})~=~$10.15\pm0.18$ and $10.14\pm0.18$ for the syntheses based on K~{\sc i} and CH, respectively. The quoted uncertainties in the Li~{\sc i} column density determinations include the usual contributions from photon noise and continuum placement errors, but include an additional contribution of 0.08 dex (added in quadrature) to account for the uncertainties involved in modeling the DIB profiles. The magnitude of this additional systematic uncertainty is based on the variation in the Li~{\sc i} column density that corresponds to a range of acceptable models for the $\lambda$6706.6 DIB (see Figure~4).

\section{RESULTS}
Table~3 provides the final column densities of the atomic and molecular species observed in the direction of SN~2014J, where we distinguish between absorption associated with the Milky Way and with M82 (taking the dividing line to be $v_{\mathrm{LSR}}$~=~+30~km~s$^{-1}$; see below). In cases where two lines from the same species were observed and fitted, the final column densities are based on the weighted means of the column densities derived from the two transitions. (This also applies to the Li~{\sc i} results based on the two independent syntheses.) We list separately the column densities of CN in the first two rotational levels, $N_{N=0}$ and $N_{N=1}$, and the total CN column density, $N_{\mathrm{tot}}$~=~$N_{N=0}$~+~$N_{N=1}$ (neglecting any contribution from the $N=2$ level). We also give the rotational excitation temperature derived from these values, $T_{01}$(CN)~=~$2.6\pm0.7$~K, which (although not very precise) is consistent with excitation by the cosmic microwave background, as found for sight lines in the local Galactic ISM (e.g., Palazzi et al.~1992; Roth \& Meyer 1995; Ritchey et al.~2011), in the Large and Small Magellanic Clouds (LMC and SMC; Welty et al.~2006), and in the host galaxies of other SNe where CN has been detected (Patat et al.~2007; Cox \& Patat 2014). Table~4 gives results for the individual absorption components discerned through the profile fitting process. No velocities are listed for Ca~{\sc i} or for the molecular species since the (relative) velocities adopted in those fits were based on the velocities found for other species (i.e., Ca~{\sc ii} for Ca~{\sc i} and CH$^+$; K~{\sc i} for CH and CN). Likewise, no $b$-values are listed for CN because the $b$-values adopted in the fits of the CN $R$(0) and $R$(1) lines were based on those found for CH.

\begin{figure}
\centering
\includegraphics[width=0.9\textwidth]{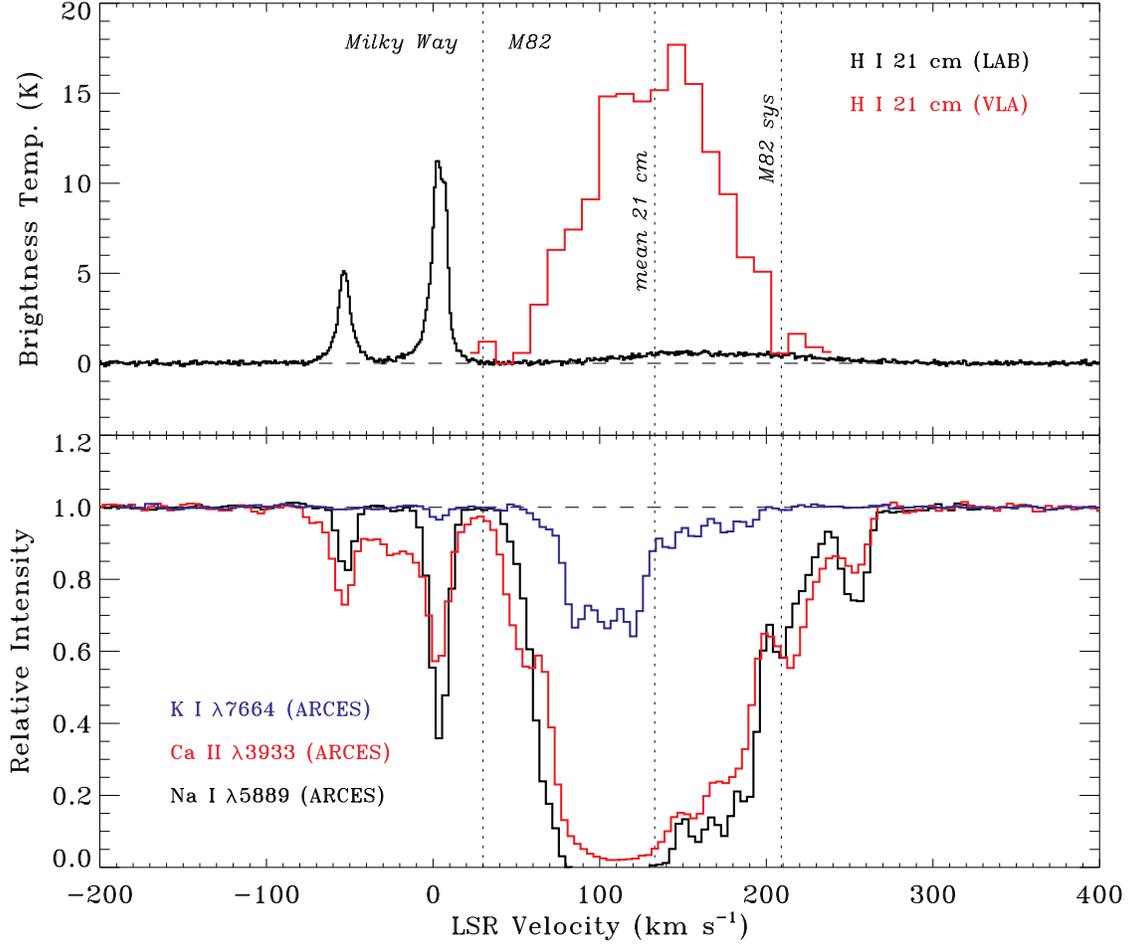}
\caption[]{Comparison between H~{\sc i}~21~cm emission profiles extracted at the position of SN~2014J and the K~{\sc i}, Ca~{\sc ii}, and Na~{\sc i} absorption profiles seen in our combined ARCES spectra. The LAB data have an angular resolution of $40\arcmin$ (Kalberla et al.~2005), while the VLA observations were obtained at a resolution of $20\arcsec$ (M.~Yun 2014, private communication). Absorption from Na~{\sc i}~$\lambda5895$ has been removed from the Na~{\sc i}~$\lambda5889$ profile shown here. The vertical dotted lines indicate the separation between Milky Way and M82 gas (at $v_{\mathrm{LSR}}$~$\approx$~+30~km~s$^{-1}$), the intensity-weighted mean H~{\sc i} velocity of M82 in the direction of SN~2014J from the VLA data ($\langle v_{\mathrm{LSR}} \rangle$~=~+133~km~s$^{-1}$), and the M82 systemic velocity ($v_{\mathrm{LSR}}$~=~+209~km~s$^{-1}$).}
\end{figure}

It is relatively straightforward to distinguish between Milky Way and M82 absorption toward SN~2014J by examining available data on H~{\sc i}~21~cm emission in that direction. Figure~5 presents a comparison between 21~cm emission profiles extracted at the position of the SN and the interstellar absorption profiles seen in our ARCES spectra. The H~{\sc i} data come from the Leiden/Argentine/Bonn (LAB) Survey (Kalberla et al.~2005; Hartmann \& Burton 1997), which provides data on Galactic H~{\sc i} emission at a resolution of $40\arcmin$ (for northern declinations), and from observations of M82 acquired at a resolution of $20\arcsec$ with the Very Large Array (VLA; M.~Yun 2014, private communication). The closest LAB sample point to SN~2014J is $8\farcm6$ away, while the closest sample point in the VLA observations is $0\farcs55$ away. The differences in pointing, therefore, are much smaller than the relevant beam sizes. The LAB and VLA data indicate that the absorption components with $v_{\mathrm{LSR}}$~$\lesssim$~+30~km~s$^{-1}$ correspond to gas within the Milky Way, while components at higher velocities are associated with material in M82. (In Section~4.1, we examine the kinematics of these absorption systems in more detail.) By integrating the LAB and VLA emission profiles over the appropriate velocity ranges (and assuming no self absorption), we obtain H~{\sc i} column densities of log~$N$(H~{\sc i})~=~20.56 and 21.46 for the gas associated with the Milky Way and M82, respectively. It is worth noting that 21~cm emission from M82 can also be seen in the LAB spectrum, but the emission is much weaker than in the VLA data because of the much larger beam (which includes the entire galaxy and much of the surrounding intergalactic medium).

The integrated H~{\sc i} column densities derived from the 21~cm emission profiles can be compared with other estimates for the hydrogen column densities based on the absorption features seen in our ARCES spectra. From the estimated Milky Way and M82 contributions to the equivalent width of the blended $\lambda$5780.5 DIB toward SN~2014J ($W_{\lambda}$(MW)~$\sim$~24 m\AA{}; $W_{\lambda}$(M82)~$\sim$~319 m\AA{}; see Welty et al.~2014), and using the relationship log~$N$(H~{\sc i})~=~19.05~+~0.92~$\times$~log~$W_{\lambda}$(5780.5) derived for sight lines in the local Galactic ISM (e.g., Friedman et al.~2011)\footnote{The coefficients defining the various correlations discussed in this section may differ slightly from those given in the quoted references since they are based on fits using a somewhat larger sample of Galactic data now available (see also Welty et al.~2014).}, we obtain log~$N$(H~{\sc i})~=~20.32 (MW) and 21.35 (M82). The CH column densities toward SN~2014J imply log~$N$(H$_2$)~=~19.93 for the Milky Way gas and 21.15 for M82, under the assumption that log~$N$(H$_2$)~=~6.82~+~1.045~$\times$~log~$N$(CH), as seen in our Galaxy and in the LMC (e.g., Welty et al.~2006). We can also estimate the total hydrogen column densities, $N$(H$_{\mathrm{tot}}$)~=~$N$(H~{\sc i})~+~$2N$(H$_2$), from the K~{\sc i} column densities, using the Galactic relationship log~$N$(K~{\sc i})~=~$-$24.325~+~1.70~$\times$~log~$N$(H$_{\mathrm{tot}}$) (e.g., Welty \& Hobbs~2001), which gives log~$N$(H$_{\mathrm{tot}}$)~=~20.49 for the Milky Way gas and 21.53 for the gas associated with M82. However, these values for $N$(H$_{\mathrm{tot}}$), when combined with the H$_2$ column densities derived from $N$(CH), would imply respective Milky Way and M82 values for log~$N$(H~{\sc i}) of 20.16 and 20.77, both of which are lower than the corresponding value predicted from the equivalent width of $\lambda$5780.5 DIB. The difference in the predictions for $N$(H~{\sc i}) is particularly severe in the case of M82, and would be even more severe if the $N$(H~{\sc i}) predicted from $W_{\lambda}$(5780.5) is slightly underestimated due to the slightly sub-solar metallicity of M82 (Origlia et al.~2004; but see Welty et al.~2014).

An important factor that may be influencing these predictions is the unusually high abundance of CH$^+$ that we find for both the Milky Way gas and the gas associated with M82. In both cases, the $N$(CH$^+$)/$N$(CH) ratio is found to be significantly larger than the values typically seen in the local Galactic ISM (see Section~4.2), suggesting that the usual relationship between CH and H$_2$ may not apply for these absorption systems, since CH can be produced via non-thermal chemistry along with CH$^+$ (e.g., Zsarg{\'o} \& Federman~2003). In principle, the H$_2$ column densities could be lower than expected by an order of magnitude or more, as seen, for example, toward 23 Ori (Welty et al.~1999) and Herschel 36 (Dahlstrom et al.~2013). The $N$(CN)/$N$(CH) ratio for the M82 material does not seem to be unusual, however, implying that a significant fraction of the CH in M82 is due to equilibrium chemistry. Still, other estimates for the H$_2$ column densities toward SN~2014J are lower than those based on $N$(CH). The K~{\sc i} column densities imply log~$N$(H$_2$)~=~19.00 for the Milky Way material and 20.91 for the gas in M82, using the Galactic relationship log~$N$(K~{\sc i})~=~$-$7.06~+~0.925~$\times$~log~$N$(H$_2$) (e.g., Welty \& Hobbs~2001). These values for $N$(H$_2$), when combined with the total hydrogen column densities predicted from $N$(K~{\sc i}), would yield Milky Way and M82 values for log~$N$(H~{\sc i}) of 20.47 and 21.25, respectively. The latter value (at least) is in better agreement with the prediction based on $W_{\lambda}$(5780.5). An additional set of estimates for $N$(H$_2$) can be obtained by combining the H~{\sc i} column densities from $W_{\lambda}$(5780.5) with the total hydrogen column densities from $N$(K~{\sc i}); these quantities yield log~$N$(H$_2$)~=~19.71 for the Milky Way and 20.76 for M82.

Altogether, we have three independent estimates each for $N$(H~{\sc i}) and $N$(H$_2$), obtained from Galactic relationships involving $W_{\lambda}$(5780.5), $N$(CH), and $N$(K~{\sc i}). The mean values (and 1$\sigma$ standard deviations) of these estimates are given in Table~3, along with the corresponding total hydrogen column densities and molecular fractions, $f$(H$_2$)~=~$2N$(H$_2$)/$N$(H$_{\mathrm{tot}}$). For the Milky Way gas, the mean predicted H~{\sc i} column density is $\sim$0.2 dex lower than the value obtained from direct integration of the LAB data, which may suggest that the $40\arcmin$ beam used with the LAB survey is sampling slightly higher-$N$(H~{\sc i}) material not directly in the SN line of sight. For the gas associated with M82, the mean of the predictions for $N$(H~{\sc i}) is $\sim$0.3 dex below the integrated H~{\sc i} column density derived from the VLA data, which could indicate that SN~2014J is probing only $\sim$50\% of the H~{\sc i} disk of M82. However, since the $20\arcsec$ beam used with the VLA observations corresponds to $\sim$340~pc at 3.5~Mpc, the differences in $N$(H~{\sc i}) could also be due to structure transverse to the line of sight, in addition to any foreground/background effects. Overall, the interstellar absorption-line tracers indicate a low to moderate molecular fraction, $f$(H$_2$)~$\approx$~$0.3\pm0.3$, for the Milky Way gas toward SN~2014J, and a moderate to high molecular fraction, $f$(H$_2$)~$\approx$~$0.5\pm0.3$, for the gas in M82.

High-resolution observations of CO emission from M82 support the notion that SN~2014J probes only about half of the gaseous disk material seen in that direction. Observations of the CO ($J=1\to0$) transition acquired at a resolution of $3\farcs5$ (F.~Walter 2014, private communication) yield an integrated intensity of $I$(CO)~$\approx$~48~K~km~s$^{-1}$ at the position of SN~2014J. Assuming an $X$-factor for M82 of $X_{\mathrm{M82}}$~=$N$(H$_2$)/$I$(CO)~=~$0.5\times10^{20}$~cm$^{-2}$~(K~km~s$^{-1}$)$^{-1}$ (Walter et al.~2002), this integrated CO intensity implies log~$N$(H$_2$)~$\approx$~21.38. The mean H$_2$ column density predicted from the absorption-line tracers toward SN~2014J is $\sim$40\% of this value. Moreover, the peak in CO emission appears to be shifted (by $\sim$+50~km~s$^{-1}$) relative to the CH and CN absorption features (and the CO profile is somewhat broader than the profiles for CH and CN). The peak in the H~{\sc i} emission profile also appears to be shifted to higher velocities relative to the atomic species seen in absorption toward the SN; the H~{\sc i} emission peaks between $v_{\mathrm{LSR}}$~$\approx$~+100 and +160~km~s$^{-1}$, while the strongest absorption components are found between $v_{\mathrm{LSR}}$~$\approx$~+80 and +125~km~s$^{-1}$ (see Figure~5). The larger column densities derived from the H~{\sc i} and CO emission data, along with the shifts in velocity, seem to indicate that a significant amount of atomic and molecular gas in M82 lies behind the SN.

\section{DISCUSSION}

\subsection{Kinematics along the Line of Sight to SN~2014J}
The dominant Milky Way absorption component at $v_{\mathrm{LSR}}$~$\approx$~+3.7~km~s$^{-1}$ (which is seen in Ca~{\sc ii}, Na~{\sc i}, K~{\sc i}, CH$^+$, and CH) is easily understood as originating in local gas in the Galactic disk. The line of sight to SN~2014J (at Galactic coordinates $l=141.4$, $b=+40.6$) passes relatively close to several of the high-latitude molecular clouds mapped in CO emission by Magnani et al.~(1985; MBM). Four of those clouds (MBM 29, 30, 31, and 32) lie within $5\degr$ of the SN position and exhibit LSR velocities between $-$0.7 and +4.0~km~s$^{-1}$. The closest (MBM 30) is only $2\fdg4$ away and has $v_{\mathrm{LSR}}$~$\approx$~+2.7~km~s$^{-1}$, in agreement with the velocity of the strongest H~{\sc i} emission component seen in the LAB data. (The strongest emission feature in the LAB profile is actually split into two subcomponents at $v_{\mathrm{LSR}}$~$\approx$~+2.8 and +6.0~km~s$^{-1}$; Figure~5). The SN line of sight is likely probing the diffuse envelope of one (or more) of these high-latitude molecular clouds, which have characteristic distances of $\sim$100~pc (Magnani et al.~1985). The velocity of the other prominent Milky Way absorption component at $v_{\mathrm{LSR}}$~$\approx$~$-$52.6~km~s$^{-1}$ (seen in Ca~{\sc ii} and Na~{\sc i}) is consistent with that predicted for co-rotating halo gas with a distance of 2.5~$\lesssim$~$d$~$\lesssim$~12.5 kpc (e.g., Morton \& Blades~1986). However, since peculiar motions could also produce such a velocity, any distance estimated in this manner will be uncertain. The other Milky Way components (seen only in Ca~{\sc ii} absorption) likely also have an origin in the Galactic disk or lower halo.

The systemic velocity of M82 (with respect to the LSR) is +209~km~s$^{-1}$ (de Vaucouleurs et al.~1991, applying a heliocentric-to-LSR velocity correction of +6~km~s$^{-1}$), and the galaxy is viewed nearly edge on (at an inclination of $80\degr$; de Vaucouleurs et al.~1991). The apparent position of SN~2014J places it $55\farcs2$ W and $20\farcs0$ S of the nucleus of M82, well within the optical disk of the galaxy. These coordinates correspond to a projected distance of 1.0~kpc from the nucleus, assuming a distance of 3.5~Mpc for M82 (Jacobs et al.~2009; Dalcanton et al.~2009). Numerous H~{\sc i} and CO emission studies have revealed unusual properties of the atomic and molecular gas in M82, which is interacting with M81 and other members of the M81 group (e.g., Yun et al.~1993, 1994; Taylor et al.~2001; Walter et al.~2002). The main features seen in H~{\sc i} and CO emission include tidal streamers emanating from the optical disk and an outflow driven by starburst activity in the central region. However, while the outer disk of M82 shows evidence of severe disruptions induced by tidal interactions with M81, rotation still dominates the kinematics of the inner 1~kpc (e.g., Yun et al.~1993; Walter et al.~2002). The sight line to SN~2014J penetrates the approaching side of the disk, but in a region characterized by a drop in the rotation curve possibly due to the presence of a tidally-induced stellar bar (Yun et al.~1993). The intensity-weighted mean H~{\sc i} velocity at the position of the SN (from the VLA data plotted in Figure~5) is $\langle v_{\mathrm{LSR}} \rangle$~=~+133~km~s$^{-1}$. Most of the absorption components associated with M82 toward SN~2014J appear to be at velocities consistent with the H~{\sc i} emission. However, (as noted above) the strongest absorption components are significantly shifted toward lower velocities compared to the peak in the H~{\sc i} profile. Furthermore, several components are found at velocities that are higher than the overall systemic velocity of M82, indicating motion in the direction opposite to that characterizing disk rotation. Extreme examples are presented by the two components seen in Ca~{\sc ii} and Na~{\sc i} absorption at $v_{\mathrm{LSR}}$~$\approx$~+249 and +258~km~s$^{-1}$. With respect to the mean H~{\sc i} velocity at the SN position, these components have velocities of +116 and +125~km~s$^{-1}$, similar to the velocities exhibited by high-velocity clouds (HVCs) in the halo of the Milky Way.

\subsection{Physical Conditions in the Milky Way and M82 Components}
The relative abundances of the various atomic and molecular species commonly studied via optical interstellar absorption lines can yield important information regarding the local physical conditions in the absorbing gas. The $N$(Na~{\sc i})/$N$(Ca~{\sc ii}) ratio is a fairly good indicator of the overall level of dust grain depletion, which ultimately depends both on the local conditions and on the grain growth/destruction history of the material. Larger $N$(Na~{\sc i})/$N$(Ca~{\sc ii}) ratios are typically found in colder, denser quiescent clouds, where the majority of Ca atoms are locked up in grains, while smaller values are characteristic of both warmer, less-shielded gas, where significant grain growth may be inhibited (e.g., Barlow 1978), and gas moving at high peculiar velocities, where dust grains are destroyed via shock processing (e.g., Siluk \& Silk~1974; Jura~1976; Shull et al.~1977). However, considering the difference in ionization potential between Na~{\sc i} and Ca~{\sc ii}, some variations in the $N$(Na~{\sc i})/$N$(Ca~{\sc ii}) ratio will be due to ionization effects rather than to changes in Ca depletion (e.g., Welty et al.~1996). In this respect, the $N$(Ca~{\sc i})/$N$(K~{\sc i}) ratio may be a better indicator of overall depletion, at least for those components where Ca~{\sc i} is detected (e.g., Welty et al.~2003). The $N$(Ca~{\sc i})/$N$(Ca~{\sc ii}) ratio for such components can then yield constraints on the ionization balance, although Ca~{\sc ii} probably occupies a somewhat larger volume than Ca~{\sc i} does in a typical diffuse interstellar cloud (Welty et al.~2003; Pan et al.~2005). Column density ratios among the molecular species CN, CH, and CH$^+$ can also help to constrain the local cloud conditions since the different molecules are typically associated with gas at different densities/temperatures, even for components at the same radial velocity (e.g., Pan et al.~2005; Welty et al.~2006; Sheffer et al.~2008). The CN molecule mainly probes the colder, denser parts of the cloud, while CH$^+$ primarily traces the warmer, more diffuse regions; CH can be associated with both CN and CH$^+$-rich gas (e.g., Federman et al.~1994; 1997).

\begin{figure}
\centering
\includegraphics[width=0.8\textwidth]{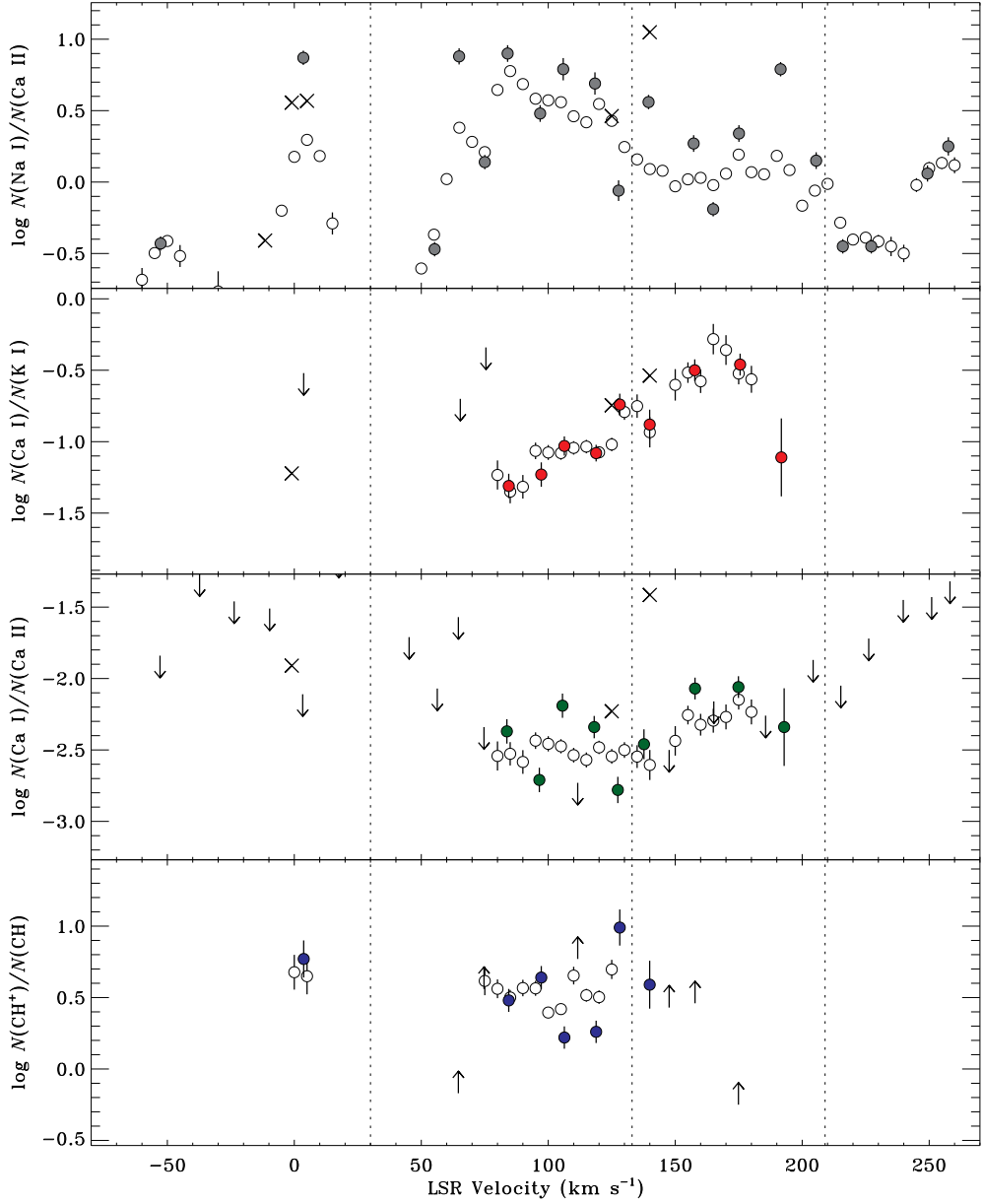}
\caption[]{Ratios of $N$(Na~{\sc i})/$N$(Ca~{\sc ii}), $N$(Ca~{\sc i})/$N$(K~{\sc i}), $N$(Ca~{\sc i})/$N$(Ca~{\sc ii}), and $N$(CH$^+$)/$N$(CH) for individual velocity components toward SN~2014J (solid colored circles). In cases where only one of the two species being plotted was detected, 3$\sigma$ upper or lower limits on the ratio were determined, as indicated. Also shown are the apparent column density ratios obtained directly from the absorption profiles via AOD calculations (open circles). (The AOD profiles were sampled at intervals of 5~km~s$^{-1}$.) Column density ratios from Vladilo et al.~(1994) for absorption components observed at $v_{\mathrm{LSR}}$~$\gtrsim$~$-$65~km~s$^{-1}$ toward SN~1993J are plotted for comparison ($\times$'s). The vertical dotted lines have the same meaning as those in Figure~5.}
\end{figure}

Since all of the above atomic and molecular species are detected toward SN~2014J, we can conduct a detailed investigation of the physical conditions in the Milky Way and M82 components (under the assumption that the various column density ratios will have the same meaning for the M82 gas as they would for gas associated with the Milky Way.) Figure~6 presents the $N$(Na~{\sc i})/$N$(Ca~{\sc ii}), $N$(Ca~{\sc i})/$N$(K~{\sc i}), $N$(Ca~{\sc i})/$N$(Ca~{\sc ii}), and $N$(CH$^+$)/$N$(CH) ratios for the individual velocity components observed in the direction of the SN (Table~4). When only one of the two species showed detectable absorption for a particular component, 3$\sigma$ upper or lower limits on the ratio were determined, as indicated in the figure. Also shown in Figure~6 are the apparent column density ratios obtained directly from the absorption profiles via AOD calculations. Several interesting trends may be noted in the figure. The general decline in the $N$(Na~{\sc i})/$N$(Ca~{\sc ii}) ratio with increasing velocity for the M82 components (particularly those with +60~$\lesssim$~$v_{\mathrm{LSR}}$~$\lesssim$~+230~km~s$^{-1}$) is matched by a steadily increasing $N$(Ca~{\sc i})/$N$(K~{\sc i}) ratio. Both of these trends indicate lower depletion for the higher-velocity components, a result which is consistent with expectations based on the Routly-Spitzer effect (Routly \& Spitzer 1952). The component at $v_{\mathrm{LSR}}$~$\approx$~+191~km~s$^{-1}$ is found to be an exception to these trends, exhibiting high $N$(Na~{\sc i})/$N$(Ca~{\sc ii}) and low $N$(Ca~{\sc i})/$N$(K~{\sc i}). The two components at $v_{\mathrm{LSR}}$~$\approx$~+249 and +258~km~s$^{-1}$ also have relatively high $N$(Na~{\sc i})/$N$(Ca~{\sc ii}) ratios compared to other components at similar velocities. The highest $N$(Na~{\sc i})/$N$(Ca~{\sc ii}) ratios exhibited by the M82 components are very similar to the ratio found for the Milky Way disk component at $v_{\mathrm{LSR}}$~$\approx$~+3.7~km~s$^{-1}$. Likewise, the lowest ratios found for gas associated with M82 are similar to the ratio determined for the Milky Way component at $v_{\mathrm{LSR}}$~$\approx$~$-$52.6~km~s$^{-1}$, which likely probes gas in the Galactic halo. There is considerably more scatter in the $N$(Na~{\sc i})/$N$(Ca~{\sc ii}) ratios than in the $N$(Ca~{\sc i})/$N$(K~{\sc i}) ratios for adjacent velocity components. This may be related to variations in the ionization conditions, since similar variations are observed in the $N$(Ca~{\sc i})/$N$(Ca~{\sc ii}) ratios. The $N$(CH$^+$)/$N$(CH) ratios also show significant scatter and appear to be anti-correlated with the $N$(Ca~{\sc i})/$N$(Ca~{\sc ii}) ratios for components where all four species are detected.

In photoionization equilibrium, the ratio $N$(Ca~{\sc i})/$N$(Ca~{\sc ii})~=~$n_e/(\Gamma/\alpha_r)$ depends on the electron density $n_e$ (which may be written as the product of the fractional ionization and the gas density), and inversely on the strength of the ambient radiation field (through the photoionization rate $\Gamma$) and the temperature (through the radiative recombination rate $\alpha_r$). For sight lines through the local Galactic ISM, typical values of log~$N$(Ca~{\sc i})/$N$(Ca~{\sc ii}) range approximately from $-$1.8 to $-$2.8 (e.g., Welty et al.~2003). All of the $N$(Ca~{\sc i})/$N$(Ca~{\sc ii}) ratios exhibited by the M82 components fall within this range. In contrast, the $N$(CH$^+$)/$N$(CH) ratios for the M82 components (and also for the Milky Way disk component) are found to be significantly larger than typical Galactic values, an indication that non-thermal chemical processes (related to the dissipation of turbulence, for instance) are highly active in these regions (e.g., Zsarg{\'o} \& Federman~2003). For the M82 components alone, the values of log~$N$(CH$^+$)/$N$(CH) range from +0.2 to +1.0, while a value of $\sim$0.0 is more typical of sight lines through the local ISM of our Galaxy (e.g., Welty et al.~2006, 2014). In environments where CH is closely associated with CH$^+$ (i.e., in gas without appreciable amounts of CN), simple chemical arguments suggest that the $N$(CH$^+$)/$N$(CH) ratio should be proportional to the strength of the UV radiation field (responsible for dissociating CH), and inversely proportional to the product of the molecular fraction and the gas density (e.g., Welty et al.~2006; Ritchey et al.~2006). The situation is more complicated for the M82 components because at least some of these do show significant column densities of CN, which is tied to CH through equilibrium chemistry (e.g., Federman et al.~1994). Still, the apparent anti-correlation between the $N$(Ca~{\sc i})/$N$(Ca~{\sc ii}) and $N$(CH$^+$)/$N$(CH) ratios can be understood in terms of the opposite dependence on gas density and radiation field strength that these two quantities are expected to exhibit. The largest $N$(CH$^+$)/$N$(CH) ratio is found for the component at $v_{\mathrm{LSR}}$~$\approx$~+128~km~s$^{-1}$, which has no associated CN absorption and exhibits the smallest $N$(Ca~{\sc i})/$N$(Ca~{\sc ii}) ratio. Conversely, the lowest $N$(CH$^+$)/$N$(CH) ratios are associated with the components at $v_{\mathrm{LSR}}$~$\approx$~+106 and +119~km~s$^{-1}$, which have the largest CN column densities and among the highest $N$(Ca~{\sc i})/$N$(Ca~{\sc ii}) ratios.

The overall high CH$^+$ abundance in the M82 gas toward SN~2014J may be indicative of enhanced turbulence in the ISM of this starburst galaxy. Recent models of turbulent dissipation regions (TDRs; Godard et al.~2009, 2014) find that the CH$^+$ abundance is directly proportional to the average turbulent dissipation rate and to the square root of the radiation field strength, and inversely proportional to the square of the gas density (see Equation 12 in Godard et al.~2014). For the M82 gas toward SN~2014J, we find a total CH$^+$ abundance of log~$N$(CH$^+$)/$N$(H$_{\mathrm{tot}}$)~$\approx$~$-$7.2 (adopting the value of $N$(H$_{\mathrm{tot}}$)$_{\mathrm{pred}}$ from Table~3). Compared to the average value for the local Galactic ISM ($\sim$$-$8.1; Godard et al.~2014), the CH$^+$ abundance in M82 seems to be enhanced by approximately 0.9~dex. Within the framework of the TDR model, a higher CH$^+$ abundance could result from a higher turbulent dissipation rate, a lower gas density, or a stronger radiation field, compared to the standard model (which adopts a gas density of $n_{\mathrm{H}}=50$~cm$^{-3}$, a visual extinction of $A_V=0.4$~mag, and a radiation field scaling factor of $\chi=1$). A low average density for the gas in M82 seems to be ruled out by the presence of CN absorption over much the same range in velocity where CH$^+$ absorption is strong. Likewise, it seems unlikely that the M82 material is immersed in a stronger (than average) radiation field. As discussed by Welty et al.~(2014), the ratio of the equivalent widths of the $\lambda5797.1$ and $\lambda5780.5$ DIBs in M82 is fairly high ($W_{\lambda}$(5797.1)/$W_{\lambda}$(5780.5)~$\sim$~0.7), which is suggestive of a relatively weak radiation field and/or a somewhat shielded environment. Moreover, the $W_{\lambda}$(5797.1)/$W_{\lambda}$(5780.5) ratio seems to be even higher (i.e., $\sim$1.1) in the stronger absorption components, with which most of the CH$^+$ is associated (Welty et al.~2014). The most likely scenario to explain the high CH$^+$ abundance within the TDR framework is, therefore, that the turbulent dissipation rate is significantly enhanced in M82. Such enhanced interstellar turbulence might be expected in a galaxy like M82, which is undergoing a starburst most likely triggered by its recent interaction with M81. The enhanced turbulence could be directly related to the interaction or could be tied to an increase in the star formation and SN rates resulting from the starburst activity.

\subsection{Comparison with the Line of Sight to SN~1993J}

High and moderately-high resolution UV and optical spectra of the Type IIb SN~1993J in M81 revealed absorption from a variety of neutral and singly-ionized atomic species over a large range in radial velocity (de Boer et al.~1993; Vladilo et al.~1993, 1994; Bowen et al.~1994; Marggraf \& de Boer~2000). Absorption from the strong Mg~{\sc ii} lines near 2800~\AA{}, for instance, extended from $v_{\mathrm{LSR}}$~=~$-$160 to +230~km~s$^{-1}$ (Bowen et al.~1994). The absorption systems detected at $-$135~$\lesssim~v_{\mathrm{LSR}}~\lesssim$~$-$115~km~s$^{-1}$ were understood as likely tracing quiescent gas in the disk of M81, while the components found at $-$63~$\lesssim~v_{\mathrm{LSR}}~\lesssim$~+18~km~s$^{-1}$ were assumed to probe gas in the Galactic disk and halo. The absorption features detected at +40~$\lesssim~v_{\mathrm{LSR}}~\lesssim$~+230~km~s$^{-1}$, however, were more difficult to interpret since it was unlikely that they belonged to either the Milky Way or M81. Several of these components exhibited relatively large $N$(Na~{\sc i})/$N$(Ca~{\sc ii}) ratios and relatively small $N$(Mg~{\sc i})/$N$(Na~{\sc i}) ratios, indicative of cool, dense gas as would normally be associated with the disk of a galaxy like the Milky Way or M81 (Vladilo et al.~1994; Bowen et al.~1994). The consensus view became that these high positive velocity components traced the remains of tidally-stripped disk material from one of the other galaxies in the M81 group, with the likely parent galaxy being M82, since it was the only major galaxy in the group with associated gaseous material at the appropriate velocity. Indeed, the high positive velocity components observed toward SN~1993J may be related to the southern H~{\sc i} streamer that Yun et al.~(1993) describe as the M81-M82 tidal tail.

The apparent position of SN~1993J in M81 (at Galactic coordinates $l=142.2$, $b=+40.9$) places it approximately $39\farcm2$ from SN~2014J in M82. A comparison between the absorption profiles for the two lines of sight reveals both similarities and differences in the components found at similar velocities (Vladilo et al.~1994; Bowen et al.~1994; Figure~3). The dominant Milky Way disk component toward SN~1993J, which in the spectra presented by Vladilo et al.~(1994) and Bowen et al.~(1994) is resolved into two subcomponents at $v_{\mathrm{LSR}}$~$\approx$~$-$1 and +5~km~s$^{-1}$, is generally stronger (in the atomic species) than the component we find at $v_{\mathrm{LSR}}$~$\approx$~+3.7~km~s$^{-1}$ toward SN~2014J, in accordance with the LAB 21~cm emission profiles in these directions (Kalberla et al.~2005; Hartmann \& Burton 1997). However, the corresponding CH and CH$^+$ components are stronger toward SN~2014J than toward SN~1993J. Both sight lines are likely probing the same complex of high-latitude molecular cloud envelopes in the Galactic disk. (At a distance of 100~pc, the two sight lines would be separated by $\sim$1.1~pc.) As seen in Figure~6, the $N$(Na~{\sc i})/$N$(Ca~{\sc ii}) ratios for all of the components near 0~km~s$^{-1}$ in the two directions are relatively high, and the $N$(Ca~{\sc i})/$N$(K~{\sc i}) ratio for the $-$1~km~s$^{-1}$ component toward SN~1993J is relatively low, suggesting a high degree of Ca depletion, as expected for cold, quiescent disk gas. The other prominent Milky Way component near $-$50~km~s$^{-1}$ is somewhat weaker in Na~{\sc i} (but stronger in Ca~{\sc ii}) toward SN~1993J than toward SN~2014J. The H~{\sc i} emission at this velocity (as seen in the LAB data) is also weaker toward the former than toward the latter. In both directions, the components near $-$50~km~s$^{-1}$ exhibit low $N$(Na~{\sc i})/$N$(Ca~{\sc ii}) ratios, as would be appropriate for halo gas with low levels of Ca depletion.

The absorption components detected at high positive velocities toward SN~1993J also show some interesting similarities, as well as differences, when compared with the M82 components toward SN~2014J. (At 3.5 Mpc, the angular separation between the two lines of sight corresponds to $\sim$40 kpc.) The $N$(Ca~{\sc i})/$N$(K~{\sc i}) ratios for the two prominent high positive velocity components toward SN~1993J, at $v_{\mathrm{LSR}}$~$\approx$~+125 and +140~km~s$^{-1}$, are very similar to the ratios we find for the M82 components at similar velocities (Figure~6). The moderately high values of $N$(Ca~{\sc i})/$N$(K~{\sc i}) for these components are indicative of relatively low levels of Ca depletion. It is perhaps somewhat surprising then that the $N$(Na~{\sc i})/$N$(Ca~{\sc ii}) ratio for the +140~km~s$^{-1}$ component toward SN~1993J is also relatively high, as high as the highest values seen among the M82 components toward SN~2014J. While this would normally be interpreted as evidence for a high degree of Ca depletion, the $N$(Ca~{\sc i})/$N$(Ca~{\sc ii}) ratio for the +140~km~s$^{-1}$ component is high as well, significantly higher than any values seen among the M82 components. The combination of a high $N$(Ca~{\sc i})/$N$(K~{\sc i}) ratio and high $N$(Na~{\sc i})/$N$(Ca~{\sc ii}) and $N$(Ca~{\sc i})/$N$(Ca~{\sc ii}) ratios suggests that the high positive velocity gas in front of M81 is located in a region characterized by a weak ambient radiation field, as might be expected for intergalactic gas situated far from any sources of ionizing photons (Vladilo et al.~1994).\footnote{The $N$(Na~{\sc i})/$N$(Ca~{\sc ii}) and $N$(Ca~{\sc i})/$N$(Ca~{\sc ii}) ratios for the +125~km~s$^{-1}$ component toward SN~1993J are probably larger than those reported by Vladilo et al.~(1994) and plotted in Figure~6. In analyzing the Ca~{\sc ii} profiles, Vladilo et al.~fit only a single broad component to the absorption at +125~km~s$^{-1}$, whereas several narrower subcomponents are probably present, as clearly seen in the Na~{\sc i} profiles. Some of this broad Ca~{\sc ii} absorption occurs at velocities outside of the range where Na~{\sc i} and Ca~{\sc i} are observed. Thus, the Ca~{\sc ii} column density will be smaller when considering only those components seen in Na~{\sc i} and Ca~{\sc i}. This will tend to make the $N$(Na~{\sc i})/$N$(Ca~{\sc ii}) and $N$(Ca~{\sc i})/$N$(Ca~{\sc ii}) ratios more consistent with those seen in the +140~km~s$^{-1}$ component.}

\subsection{The Lithium Abundance in the ISM of M82}

Our detection of Li~{\sc i} absorption over a range in velocity consistent with that exhibited by the strongest Na~{\sc i} and K~{\sc i} components associated with M82 (Figure~4) constitutes the first detection of Li in the interstellar medium of a galaxy beyond the Local Group. The only other detection of interstellar Li outside of the Milky Way is that reported by Howk et al.~(2012), who presented high-resolution observations of the Li~{\sc i} doublet along the line of sight to the star Sk~143 in the SMC. Those observations were obtained in order to address the well-known discrepancy between the primordial $^7$Li abundance predicted by standard big bang nucleosynthesis and the $^7$Li abundance observed in the atmospheres of metal-poor Galactic halo stars (e.g., Fields 2011; Spite et al.~2012). Determinations of the Li abundance in interstellar environments with sub-solar metallicities provide a method of constraining the cosmic evolution of Li that is unaffected by the uncertainties that have traditionally hampered stellar Li abundance studies. With a metallicity of [Fe/H]~$\approx$~$-$0.59 (see Howk et al.~2012), the SMC is the first low-metallicity interstellar environment where the abundance of Li has been reliably measured; Howk et al.~(2012) find a $^7$Li abundance of $A$($^7$Li)~$\equiv$~log~$N$($^7$Li)/$N$(H$_{\mathrm{tot}}$)~+~12~=~$2.68\pm0.16$ for the SMC gas toward Sk~143. For M82, Origlia et al.~(2004) find a metallicity of [Fe/H]~$\approx$~$-$0.35, from a combined analysis of integrated near-infrared stellar spectra and X-ray spectra of hot gas in the nucleus. As such, our observations of Li~{\sc i} absorption in the ISM of M82 toward SN~2014J allow us to investigate the Li abundance in an interstellar environment with a metallicity intermediate between that of the SMC and the solar system.

The derivation of the Li abundance from observations of interstellar Li~{\sc i} is complicated by the combined effects of ionization and dust grain depletion. In neutral, diffuse clouds, the dominant ionization state of Li is Li~{\sc ii}, which is unobservable, and the use of ionization corrections, based on observations of different atoms in successive ionization states for instance, often leads to conflicting and/or uncertain results (e.g., Welty et al.~2003). Consideration of the $N$(Li~{\sc i})/$N$(K~{\sc i}) ratio as a proxy for the Li abundance, however, can help to mitigate these uncertainties since Li~{\sc i} and K~{\sc i} respond to changes in the ionization and depletion conditions in a similar way (e.g., White~1986; Welty \& Hobbs~2001; Knauth et al.~2003). For the interstellar gas in M82 toward SN~2014J, our measured value of log~$N$(Li~{\sc i})/$N$(K~{\sc i})~=~$-$$2.06\pm0.18$ (for the six K~{\sc i} components used to synthesize the Li~{\sc i} feature) implies log~$N$(Li)/$N$(K)~=~$-$$1.52\pm0.20$, applying a differential ionization correction of log~$[(\Gamma/\alpha_r)_{\mathrm{Li}}/(\Gamma/\alpha_r)_{\mathrm{K}}]$~=~$+$$0.54\pm0.08$ (Steigman 1996; Welty et al.~2003). This ionization correction assumes a temperature of $T=100$~K, although the ratio of the recombination coefficients is not very sensitive to $T$ for the species we are considering (e.g., P{\'e}quignot \& Aldrovandi 1986). Comparing the above Li/K ratio to the solar system ratio of log~(Li/K)$_{\sun}$~=~$-$$1.82\pm0.05$ (Asplund et al.~2009) yields [Li/K]~$\equiv$~log~$N$(Li)/$N$(K)~$-$~log~(Li/K)$_{\sun}$~=~$+$$0.30\pm0.20$, which is suggestive of a mild Li enhancement, relative to K, for the ISM of M82. This ratio can be converted to an absolute Li abundance through use of the equation $A$(Li)$_{\mathrm{M82}}$~=~$A$(Li)$_{\sun}$~+~[K/H]$_{\mathrm{M82}}$~+~[Li/K]$_{\mathrm{M82}}$, for which we adopt a solar Li abundance of $A$(Li)$_{\sun}$~=~$3.26\pm0.05$ (Asplund et al.~2009). Since K is a product of explosive oxygen burning in massive stars and tends to behave like an $\alpha$ element (e.g., Woosley \& Weaver~1995; Samland~1998), we can use the $\alpha$ element abundances derived for M82 as a proxy for [K/H]$_{\mathrm{M82}}$. Origlia et al.~(2004) find an average $\alpha$ element abundance of [$\langle$Si,~Mg,~Ca$\rangle$/H]~$\approx$~$+$$0.04\pm0.24$ from integrated near-infrared spectra of red supergiants in the nucleus of M82.\footnote{This corresponds to an average $\alpha$ element enhancement factor of [$\alpha$/Fe]~$\approx$~$+$$0.4\pm0.3$ dex, which is consistent with a chemical evolution scenario in which the ISM of M82 is enriched by the products of Type II SNe happening in recurrent bursts on short timescales (Origlia et al.~2004).} Substituting this value for [K/H]$_{\mathrm{M82}}$ in the above equation gives $A$(Li)$_{\mathrm{M82}}$~=~$3.60\pm0.31$, where the uncertainties reflect the propagated errors from all of the quantities contributing to this determination.

In thin disk stars of the Milky Way, the Li abundance is found to increase steadily with metallicity, from a value of $A$(Li)~$\approx$~2.2 at [Fe/H]~$\approx$~$-$1.0 to a value consistent with the solar abundance at [Fe/H]~$\approx$~0.0 (Lambert \& Reddy~2004). The rise in the Li abundance with increasing metallicity (for [Fe/H]~$\gtrsim$~$-$1.0) presumably reflects the combined influence of cosmic-ray and stellar nucleosynthesis, whose relative contributions to the cosmic Li abundance also increase with metallicity (e.g., Prantzos~2012). The low Li abundance derived by Howk et al.~(2012) for interstellar gas in the SMC is consistent (within the uncertainties) with the mean Li abundance determined for Galactic thin disk stars at [Fe/H]~$\approx$~$-$0.6 (Lambert \& Reddy~2004). Our determination of $A$(Li)~=~$3.60\pm0.31$ for M82, however, is considerably higher than the Li abundance expected at a metallicity of [Fe/H]~$\approx$~$-$0.35; the mean Li abundance found by Lambert \& Reddy~(2004) at this metallicity is $A$(Li)~=~$2.86\pm0.14$, suggesting that Li could be enhanced by up to 0.7 dex in M82. While the uncertainties involved in this determination are relatively large, there are reasons that one might expect a Li enhancement in a starburst galaxy. The high star formation and SN rates in such a galaxy should lead to enhanced cosmic-ray acceleration, while the high gas density should result in frequent interactions between accelerated particles and interstellar nuclei. Such interactions would be expected to lead to an enhancement in the production of Li by cosmic-ray spallation and by $\alpha+\alpha$ fusion reactions (e.g., Ramaty et al.~1997; Lemoine et al.~1998).

Another consequence of the interaction between cosmic rays and interstellar gas is the emission of gamma rays produced by the decay of neutral pions, and starburst galaxies (M82 in particular) have long been predicted to be luminous sources of gamma-ray emission (e.g., V{\"o}lk et al.~1996). Recent observations at GeV and TeV energies have finally succeeded in detecting high-energy gamma-ray emission from M82 (Acciari et al.~2009; Abdo et al.~2010). By comparing their TeV observations, acquired with the Very Energetic Radiation Imaging Telescope Array System (VERITAS), to theoretical models, Acciari et al.~(2009) estimate that the cosmic-ray energy density in the starburst core of M82 is approximately 500 times the average Milky Way density. Abdo et al.~(2010), examining data from the Large Area Telescope (LAT) of the \emph{Fermi Gamma-ray Space Telescope} for M82 and another starburst galaxy, along with similar data for the Milky Way and the LMC, found a clear positive correlation between gamma-ray luminosity and the product of the SN rate and the total gas mass. Together, these results provide compelling evidence for a connection between star formation activity and cosmic-ray acceleration in star-forming galaxies. Our detection of an enhanced Li abundance in the ISM of M82 supports the conclusions of the gamma-ray emission studies.

\section{SUMMARY AND CONCLUSIONS}
We have presented the results of a comprehensive analysis of atomic and molecular interstellar absorption lines seen in moderately-high resolution, high S/N optical spectra of SN~2014J in M82. Our multi-epoch observations, obtained with the ARC echelle spectrograph at APO, cover the period from $\sim$6 days before to $\sim$30 days after the supernova reached its maximum $B$-band brightness on 2014 February 2. The spectra reveal complex interstellar absorption from Na~{\sc i}, Ca~{\sc ii}, K~{\sc i}, Ca~{\sc i}, CH$^+$, CH, and CN, much of which arises from gas in the ISM of M82 (at velocities of +45~$\lesssim~v_{\mathrm{LSR}}~\lesssim$~+260~km~s$^{-1}$), although absorption features associated with the Galactic disk and halo are also observed (near $v_{\mathrm{LSR}}$~$\approx$~+4 and $-$53~km~s$^{-1}$, respectively). We detect Li~{\sc i} absorption over a range in velocity consistent with that exhibited by the strongest Na~{\sc i} and K~{\sc i} components associated with M82. To our knowledge, this is the first detection of interstellar Li in a galaxy outside of the Local Group. There are no significant temporal variations in the absorption-line profiles over the 37 days sampled by our observations. We place stringent (3$\sigma$) upper limits on the column densities of any (potentially variable) Na~{\sc i}, Ca~{\sc ii}, or K~{\sc i} components not detected in the nightly sum spectra: log~$N$(Na~{\sc i})~$\lesssim$~10.3, log~$N$(Ca~{\sc ii})~$\lesssim$~11.0, and log~$N$(K~{\sc i})~$\lesssim$~10.2.

The relative abundances of the atomic and molecular species observed in the direction of SN~2014J offer insight into the local physical conditions in the absorbing gas. For the M82 components (particularly those with +60~$\lesssim$~$v_{\mathrm{LSR}}$~$\lesssim$~+230~km~s$^{-1}$), there is a general decline in the $N$(Na~{\sc i})/$N$(Ca~{\sc ii}) ratio with increasing velocity, which is matched by a steadily increasing $N$(Ca~{\sc i})/$N$(K~{\sc i}) ratio. Taken together, these trends are an indication of reduced depletion in the higher-velocity material (with a notable exception seen for the component at $v_{\mathrm{LSR}}$~$\approx$~+191~km~s$^{-1}$). We find considerably more scatter in the $N$(Na~{\sc i})/$N$(Ca~{\sc ii}) ratios than in the $N$(Ca~{\sc i})/$N$(K~{\sc i}) ratios for adjacent velocity components. This may be related to variations in the ionization conditions, since similar variations are observed in the $N$(Ca~{\sc i})/$N$(Ca~{\sc ii}) ratios. The $N$(CH$^+$)/$N$(CH) ratios also show significant scatter and appear to be anti-correlated with the $N$(Ca~{\sc i})/$N$(Ca~{\sc ii}) ratios for components where all four species are detected. The apparent anti-correlation can be understood in terms of the opposite dependence on gas density and radiation field strength that these two quantities are expected to exhibit. In general, we find that the ISM of M82 probed by SN~2014J is complex, consisting of a mixture of diffuse atomic and molecular clouds characterized by a wide range of physical/environmental conditions. The high overall abundance of CH$^+$ in M82 may be indicative of enhanced turbulence in the ISM of this starburst galaxy.

A comparison between the lines of sight to SN~1993J in M81 and SN~2014J in M82 reveals both similarities and differences in the Galactic and extragalactic components found at similar velocities. The Milky Way disk component is generally stronger (in the atomic species) toward SN~1993J than toward SN~2014J, yet the CH and CH$^+$ column densities are higher in the latter direction. This may be related to the fact that the $N$(CH$^+$)/$N$(CH) ratio exhibited by the Milky Way disk component toward SN~2014J is also quite high. The extragalactic components at high positive velocities toward SN~1993J, like the M82 components toward SN~2014J at similar velocities, have relatively high $N$(Ca~{\sc i})/$N$(K~{\sc i}) ratios, indicative of relatively low levels of Ca depletion. Yet, the components toward SN~1993J also have elevated $N$(Na~{\sc i})/$N$(Ca~{\sc ii}) and $N$(Ca~{\sc i})/$N$(Ca~{\sc ii}) ratios, suggesting that the high positive velocity gas in front of M81 is located in a region characterized by a weak ambient radiation field, as might be expected for intergalactic gas without an accompanying population of hot stars. A similar phenomenon may be responsible for the relatively high $N$(Na~{\sc i})/$N$(Ca~{\sc ii}) ratios found for the M82 components at $v_{\mathrm{LSR}}$~$\approx$~+249 and +258~km~s$^{-1}$ toward SN~2014J. Kinematic arguments suggest that these components are probing high velocity material moving in a direction opposite to that characterizing disk rotation. The velocities themselves are similar to the velocities exhibited by the western H~{\sc i} streamer mapped by Yun et al.~(1993), which terminates close to the disk of M82 near the position of SN~2014J.

Our detection of Li~{\sc i} absorption associated with interstellar gas in M82 has allowed us to examine the Li abundance in the ISM of this starburst galaxy. We find [Li/K]$_{\mathrm{M82}}$~=~$+$$0.30\pm0.20$, suggestive of a mild enhancement, relative to the solar system, in the Li/K abundance ratio for the ISM of M82. From this quantity, we infer an absolute Li abundance of $A$(Li)$_{\mathrm{M82}}$~=~$3.60\pm0.31$, which is considerably higher than the Li abundance expected at a metallicity for M82 of [Fe/H]~$\approx$~$-$0.35. Compared to the mean Li abundance found for Galactic thin disk stars at this metallicity, the Li abundance in M82 could be enhanced by up to 0.7 dex. Such an enhancement would not be unexpected for a starburst galaxy, where the combination of a high SN rate and a high gas density should lead to enhanced cosmic-ray acceleration and frequent interactions between accelerated particles and interstellar nuclei. M82 is one of only two starburst galaxies that have been detected in gamma-ray emission at GeV and TeV energies. The gamma-ray emission data suggest that the cosmic-ray density in the starburst core of M82 is approximately 500 times larger than the average density of cosmic rays in the Milky Way. Our detection of an enhanced Li abundance in the ISM of M82 supports the connection between star formation activity and cosmic-ray acceleration in star-forming galaxies that is inferred from the gamma-ray emission studies.

\acknowledgments
We are grateful to Min Yun for providing us with the H~{\sc i} emission spectrum from the VLA and to Fabian Walter for supplying the CO emission data. Support for this work has been provided by the Kenilworth Fund of the New York Community Trust (A.M.R.), and by the National Science Foundation, under grants AST-1009603 (D.G.Y.), AST-1008424 (J.D.), and AST-1238926 (D.E.W.).

{\it Facility:} \facility{ARC (ARCES)}

\clearpage

\begin{deluxetable}{lcccccc}
\tablecolumns{7}
\tablewidth{0pc}
\tabletypesize{\small}
\tablecaption{Observational Data and Single-epoch Equivalent Widths}
\tablehead{ \colhead{UT Date} & \colhead{Phase\tablenotemark{a}} & \colhead{Exp.~Time} & \colhead{S/N\tablenotemark{b}} & \colhead{$W_{\lambda}$(3968)\tablenotemark{c}} & \colhead{$W_{\lambda}$(5895)\tablenotemark{c}} & \colhead{$W_{\lambda}$(7698)\tablenotemark{c}} \\
\colhead{} & \colhead{(days)} & \colhead{(s)} & \colhead{} & \colhead{(m\AA)} & \colhead{(m\AA)} & \colhead{(m\AA)} }
\startdata
2014 Jan 27.4 & $-$5.6 & 8400 & 100/270/410 & $1394\pm28$ & $2529\pm14$ & $287.5\pm7.3$ \\
2014 Jan 30.2 & $-$2.8 & 7200 & 90/290/380 & $1377\pm32$ & $2534\pm13$ & $305.7\pm7.9$ \\
2014 Feb 8.2 & +6.2 & 12000 & 90/280/320 & $1394\pm29$ & $2532\pm14$ & $303.3\pm9.5$ \\
2014 Feb 11.3 & +9.3 & 6380 & 80/230/280 & $1387\pm33$ & $2547\pm16$ & \phn$293.1\pm11.3$ \\
2014 Feb 22.4 & +20.4 & 15300 & 20/240/370 & \ldots\tablenotemark{d} & $2544\pm16$ & $299.8\pm8.2$ \\
2014 Mar 4.4 & +30.4 & 6000 & 20/230/410 & \ldots\tablenotemark{d} & $2546\pm16$ & $301.2\pm7.4$ \\
\enddata
\tablenotetext{a}{Relative to $B$-band maximum, assumed to be 2014 February 2.0 (Foley et al.~2014).}
\tablenotetext{b}{S/N per resolution element for spectral region near Ca~{\sc ii} $\lambda3968$/Na~{\sc i} $\lambda5895$/K~{\sc i} $\lambda7698$.}
\tablenotetext{c}{Single-epoch equivalent widths for absorption associated with M82 (see Section 3).}
\tablenotetext{d}{Measurements of $W_{\lambda}$(3968) are not reported due to the very low S/N achieved.}
\end{deluxetable}

\begin{deluxetable}{lccccc}
\tablecolumns{6}
\tablewidth{0pc}
\tabletypesize{\small}
\tablecaption{Equivalent Widths from Co-added Spectra}
\tablehead{ \colhead{Species} & \colhead{Line} & \colhead{$\lambda$\tablenotemark{a}} & \colhead{$f$\tablenotemark{a}} & \colhead{$W_{\lambda}$(MW)} & \colhead{$W_{\lambda}$(M82)} \\
\colhead{} & \colhead{} & \colhead{(\AA)} & \colhead{} & \colhead{(m\AA)} & \colhead{(m\AA)} }
\startdata
Ca~{\sc ii} & K & 3933.661 & 0.6267 & $206.5\pm3.9$ & $1801\pm15$ \\
 & H & 3968.467 & 0.3116 & $116.3\pm4.1$ & $1391\pm16$ \\
Ca~{\sc i} & & 4226.728 & 1.77\phn\phn & \ldots & \phn$51.3\pm5.0$ \\
Na~{\sc i} & D$_2$ & 5889.951 & 0.6408 & $203.3\pm1.4$ & $2842\pm11$ \\
 & D$_1$ & 5895.924 & 0.3201 & $169.8\pm1.4$ & $2532\pm10$ \\
Li~{\sc i} & & 6707.826 & 0.7470 & \ldots & \phn\phn$4.1\pm2.1$ \\
K~{\sc i} & & 7664.899 & 0.6682 & \phn$10.0\pm1.0$ & $542.5\pm9.2$ \\
 & & 7698.965 & 0.3327 & \phn\phn$7.0\pm0.9$ & $298.7\pm6.0$ \\
CN & $R$(0) & 3874.608 & 0.0342 & \ldots & \phn$31.2\pm6.7$ \\
 & $R$(1) & 3873.998 & 0.0228 & \ldots & \phn\phn$8.2\pm3.8$ \\
CH$^+$ & (0, 0) & 4232.548 & \phn0.00545 & \phn$16.2\pm1.0$ & $169.2\pm6.5$ \\
 & (1, 0) & 3957.692 & \phn0.00331 & \phn\phn$9.2\pm1.5$ & \phn$93.4\pm9.0$ \\
CH & $A$$-$$X$ & 4300.313 & \phn0.00506 & \phn\phn$2.8\pm0.9$ & \phn$40.5\pm5.3$ \\
 & $B$$-$$X$ & 3886.409 & \phn0.00320 & \ldots & \phn$11.4\pm5.0$ \\
\enddata
\tablenotetext{a}{Wavelengths and oscillators strengths are from Morton (2003) for atomic lines, Roth \& Meyer (1995) for CN, and Gredel et al.~(1993) for CH$^+$ and CH.}
\end{deluxetable}

\begin{deluxetable}{lccc}
\tablecolumns{4}
\tablewidth{0pc}
\tabletypesize{\small}
\tablecaption{Column Densities and Other Sight Line Parameters}
\tablehead{ \colhead{Quantity} & \colhead{Milky Way} & \colhead{M82} & \colhead{Note} }
\startdata
log $N$(Ca~{\sc ii}) & $12.48\pm0.02$ & $13.83\pm0.05$ & 1 \\
log $N$(Na~{\sc i}) & $12.99\pm0.03$ & $14.25\pm0.04$ & 1 \\
log $N$(K~{\sc i}) & $10.52\pm0.08$ & $12.28\pm0.02$ & 1 \\
log $N$(Ca~{\sc i}) & \ldots & $11.27\pm0.04$ & 1 \\
log $N$(Li~{\sc i}) & \ldots & $10.14\pm0.18$ & 1 \\
log $N$(CH$^+$) & $13.31\pm0.02$ & $14.33\pm0.02$ & 1 \\
log $N$(CH) & $12.54\pm0.12$ & $13.71\pm0.05$ & 1 \\
log $N_{\mathrm{tot}}$(CN) & \ldots & $13.00\pm0.08$ & 1 \\
log $N_{N=0}$(CN) & \ldots & $12.86\pm0.08$ & 1 \\
log $N_{N=1}$(CN) & \ldots & $12.44\pm0.17$ & 1 \\
$T_{01}$(CN) (K) & \ldots & $2.6\pm0.7$ & 1 \\
\hline
log $N$(H~{\sc i})$_{\mathrm{21cm}}$ & $20.56\pm0.02$ & $21.46\pm0.05$ & 2 \\
log $N$(H~{\sc i})$_{\mathrm{pred}}$ & $20.33\pm0.13$ & $21.19\pm0.19$ & 3 \\
log $N$(H$_2$)$_{\mathrm{pred}}$ & $19.69\pm0.25$ & $20.97\pm0.17$ & 3 \\
log $N$(H$_{\mathrm{tot}}$)$_{\mathrm{pred}}$ & $20.49\pm0.13$ & $21.53\pm0.13$ & 4 \\
log $f$(H$_2$)$_{\mathrm{pred}}$ & $-$$0.51\pm0.26$ & $-$$0.26\pm0.20$ & 4 \\
$E(\bv)$ & 0.054 & \phn$1.2\pm0.1$ & 5 \\
\enddata
\tablecomments{(1) Obtained through profile synthesis fits to ARCES spectra; (2) Obtained by integrating the 21~cm emission profile from the LAB survey (MW; Kalberla et al.~2005) or from VLA observations (M82; M. Yun 2014, private communication); (3) Derived by averaging the values predicted by various Galactic relationships involving $W_{\lambda}$(5780.5), $N$(CH), and $N$(K~{\sc i}) (see the text); (4) Derived from $N$(H~{\sc i})$_{\mathrm{pred}}$ and $N$(H$_2$)$_{\mathrm{pred}}$; (5) MW: See Foley et al.~(2014), and references therein; M82: Mean of the values given by Goobar et al.~(2014), Amanullah et al.~(2014), and Foley et al.~(2014).}
\end{deluxetable}

\begin{deluxetable}{cccccccccccccccc}
\rotate
\tablecolumns{16}
\tablewidth{0pc}
\tabletypesize{\scriptsize}
\tablecaption{Component Structure}
\tablehead{ \colhead{$v$(Ca~{\sc ii})} & \colhead{$N$(Ca~{\sc ii})} & \colhead{$b$(Ca~{\sc ii})} & \colhead{$v$(Na~{\sc i})} & \colhead{$N$(Na~{\sc i})} & \colhead{$b$(Na~{\sc i})} & \colhead{$v$(K~{\sc i})} & \colhead{$N$(K~{\sc i})} & \colhead{$b$(K~{\sc i})} & \colhead{$N$(Ca~{\sc i})} & \colhead{$b$(Ca~{\sc i})} & \colhead{$N$(CH$^+$)} & \colhead{$b$(CH$^+$)} & \colhead{$N$(CH)} & \colhead{$b$(CH)} & \colhead{$N$(CN)} }
\startdata
\multicolumn{16}{c}{Milky Way} \\
\hline
$-$64.3 &   1.3 &  5.7  & \ldots &  \ldots & \ldots & \ldots & \ldots & \ldots & \ldots & \ldots & \ldots & \ldots & \ldots & \ldots & \ldots \\
$-$52.9 &   6.9 &  3.4  &$-$52.4 &    2.6  &  (2.0) & \ldots & \ldots & \ldots & \ldots & \ldots & \ldots & \ldots & \ldots & \ldots & \ldots \\
$-$37.3 &   1.9 &  3.5  & \ldots &  \ldots & \ldots & \ldots & \ldots & \ldots & \ldots & \ldots & \ldots & \ldots & \ldots & \ldots & \ldots \\
$-$23.7 &   2.9 &  4.4  & \ldots &  \ldots & \ldots & \ldots & \ldots & \ldots & \ldots & \ldots & \ldots & \ldots & \ldots & \ldots & \ldots \\
 $-$9.7 &   3.2 &  4.2  & \ldots &  \ldots & \ldots & \ldots & \ldots & \ldots & \ldots & \ldots & \ldots & \ldots & \ldots & \ldots & \ldots \\
   +3.3 &  13.0 &  4.0  &   +3.8 &   94.9  &  (2.0) &   +4.1 &   3.3  &   2.5  & \ldots & \ldots &  20.4  &  (3.0) &   3.5  &  (2.0) & \ldots \\
  +17.5 &   1.4 & (6.0) & \ldots &  \ldots & \ldots & \ldots & \ldots & \ldots & \ldots & \ldots & \ldots & \ldots & \ldots & \ldots & \ldots \\
\hline
\multicolumn{16}{c}{M82} \\
\hline
  +45.2 &   5.2 &  5.7  & \ldots &  \ldots & \ldots & \ldots & \ldots & \ldots & \ldots & \ldots & \ldots & \ldots & \ldots & \ldots & \ldots \\
  +56.3 &  11.9 & (6.0) &  +54.2 &    4.0  &  (6.0) & \ldots & \ldots & \ldots & \ldots & \ldots & \ldots & \ldots & \ldots & \ldots & \ldots \\
  +64.6 &   3.7 &  5.2  &  +65.2 &   28.5  &   3.0  &  +66.1 &   5.0  &   4.5  & \ldots & \ldots &   2.2  &  (6.0) & \ldots & \ldots & \ldots \\
  +74.7 &  22.0 &  4.4  &  +75.3 &   30.0  &   3.1  &  +76.2 &   2.2  &  (5.0) & \ldots & \ldots &  11.4  &  (3.0) & \ldots & \ldots & \ldots \\
  +83.7 &  46.1 &  5.3  &  +84.3 & (360.6) &   5.0  &  +85.2 &  39.9  &   4.4  &   2.0  &  (6.0) &  34.7  &   4.5  &  11.5  &  (5.0) &  (2.6) \\
  +96.5 &  85.5 & (6.0) &  +97.1 & (254.1) &   5.3  &  +98.0 &  28.4  &   4.9  &   1.7  &  (2.0) &  37.1  &   5.7  &   8.5  &  (2.0) &  (1.9) \\
 +105.6 &  41.2 & (3.0) & +106.2 & (259.2) &   4.4  & +107.1 &  28.9  &   2.4  &   2.7  &   4.4  &  21.2  &   5.7  &  12.8  &  (2.0) &  (2.9) \\
 +111.6 &  53.6 &  4.5  & \ldots &  \ldots & \ldots & \ldots & \ldots & \ldots & \ldots & \ldots &  18.5  &   4.0  & \ldots & \ldots & \ldots \\
 +118.1 &  78.2 & (3.5) & +118.7 & (381.9) &   4.0  & +119.6 &  42.8  &   4.3  &   3.6  &  (6.0) &  21.2  &   4.4  &  11.8  &  (2.0) &  (2.6) \\
 +127.4 & 119.8 & (3.5) & +128.0 &  105.2  &   4.9  & +128.9 &  10.9  &   4.6  &   2.0  &   3.4  &  35.7  &   3.6  &   3.6  &  (2.0) & \ldots \\
 +137.7 &  40.3 &  3.8  & +141.3 &  146.4  &   4.0  & +142.2 &  10.8  &  (2.0) &   1.4  &   4.1  &  11.3  &   3.2  &   2.9  &   4.0  & \ldots \\
 +147.6 &  31.8 &  5.4  & \ldots &  \ldots & \ldots & \ldots & \ldots & \ldots & \ldots & \ldots &   8.4  &   5.2  & \ldots & \ldots & \ldots \\
 +157.8 &  30.7 &  4.0  & +156.7 &   57.1  &   4.4  & +157.6 &   8.4  &   3.6  &   2.6  &   5.8  &   9.1  &   4.6  & \ldots & \ldots & \ldots \\
 +165.2 &  14.3 &  3.6  & +164.7 &    9.3  &   2.6  & \ldots & \ldots & \ldots & \ldots & \ldots & \ldots & \ldots & \ldots & \ldots & \ldots \\
 +174.9 &  28.5 &  5.4  & +175.3 &   61.1  &  (4.8) & +176.2 &   7.0  &   3.2  &   2.4  &  (6.0) &   1.8  &   5.7  & \ldots & \ldots & \ldots \\
 +185.6 &  18.2 &  4.2  & \ldots &  \ldots & \ldots & \ldots & \ldots & \ldots & \ldots & \ldots & \ldots & \ldots & \ldots & \ldots & \ldots \\
 +192.9 &   7.4 &  4.0  & +190.1 &   45.3  &   3.2  & +190.7 &   4.4  &  (2.0) &   0.3  &  (2.0) & \ldots & \ldots & \ldots & \ldots & \ldots \\
 +204.3 &   7.4 &  3.5  & +206.7 &   10.6  &  (2.0) & +206.7 &   0.3  &   2.2  & \ldots & \ldots & \ldots & \ldots & \ldots & \ldots & \ldots \\
 +215.2 &  11.1 &  4.7  & +216.8 &    4.0  &   5.6  & +217.3 &   0.2  &  (5.0) & \ldots & \ldots & \ldots & \ldots & \ldots & \ldots & \ldots \\
 +226.2 &   5.3 &  5.0  & +228.2 &    1.9  &   5.6  & \ldots & \ldots & \ldots & \ldots & \ldots & \ldots & \ldots & \ldots & \ldots & \ldots \\
 +239.8 &   2.8 &  4.4  & \ldots &  \ldots & \ldots & \ldots & \ldots & \ldots & \ldots & \ldots & \ldots & \ldots & \ldots & \ldots & \ldots \\
 +251.0 &   2.7 &  3.3  & +247.7 &    3.1  &   2.6  & \ldots & \ldots & \ldots & \ldots & \ldots & \ldots & \ldots & \ldots & \ldots & \ldots \\
 +258.2 &   2.1 &  3.2  & +256.9 &    3.8  &  (2.0) & \ldots & \ldots & \ldots & \ldots & \ldots & \ldots & \ldots & \ldots & \ldots & \ldots \\
\enddata
\tablecomments{Units for $v$ and $b$ are km~s$^{-1}$. Units for $N$ are cm$^{-2}$, where $N$(Ca~{\sc ii}) and $N$(Na~{\sc i}) must be multiplied by $10^{11}$, $N$(K~{\sc i}) and $N$(Ca~{\sc i}) by $10^{10}$, and $N$(CH$^+$), $N$(CH), and $N$(CN) by $10^{12}$. Quantities in parentheses were constrained during profile synthesis. The column densities of the four strongly-saturated Na~{\sc i} components were held fixed, assuming $N$(Na~{\sc i})/$N$(K~{\sc i})~$\approx$~90. The fractional column densities of the CN components are based on those found for the corresponding components in CH. Any $b$-values that were held fixed or that reached the edge of their allowed range are also identified with parentheses.}
\end{deluxetable}

\end{document}